%% file: main.tex
\documentclass[
    twocolumn,
    10pt,
    aps,
    pra,
    superscriptaddress,
    floatfix,
    citeautoscript
]{revtex4-2}

\usepackage[T1]{fontenc}
\usepackage[dvipsnames]{xcolor}
\usepackage{amsthm}
\usepackage{thmtools}
\usepackage{mathtools}
\usepackage{braket}
\usepackage{bm} % bold math
\usepackage{bbm} % black board math
\usepackage{dsfont}
\usepackage{amsmath}
\usepackage{amssymb}
\usepackage{enumitem}
\usepackage[colorlinks]{hyperref}
\usepackage{inputenc}
\usepackage{tikz}
\usepackage{xurl}

\usepackage{etoolbox}
\apptocmd{\sloppy}{\hbadness 10000\relax}{}{}

\usepackage{silence}
        \DeclareMathOperator{\range}{ran} % range
        \newcommand*{\abs}[1]{\lvert #1\rvert} % absolute value
        \newcommand*{\one}{\mathds{1}} % identity matrix/operator
        \newcommand*{\N}{\mathbb{N}} % natural numbers
        \newcommand*{\Z}{\mathbb{Z}} % integers
        \newcommand*{\C}{\mathbb{C}} % complex numbers
        \newcommand*{\hil}{\mathcal{H}} % Hilbert space
        \newcommand*{\lo}{\mathcal{L}} % linear bounded operators
        \newcommand*{\unitary}{\mathcal{U}} % unitary operators
        \newcommand*{\defdot}{\,\cdot\,} % dot for definitions
        \newcommand*{\defcolon}{\,:\,} % colon for e.g. definition in sets
        \newcommand\scriptin{\raisebox{0.15ex}{$\scriptscriptstyle\in$}} % "element of" symbol for usage in sub- and superscripts
        
        \newcommand*{\ketbra}[2]{\ket{#1}\negthickspace\bra{#2}} % ket-bra notation
        \newcommand*{\up}{\ket{\uparrow}} % up-arrow ket
        \newcommand*{\down}{\ket{\downarrow}} % down-arrow ket
        \newcommand*{\upup}{\ketbra{\uparrow}{\uparrow}} % up-up ketbra
        \newcommand*{\downup}{\ketbra{\downarrow}{\uparrow}} % down-up ketbra
        \newcommand*{\updown}{\ketbra{\uparrow}{\downarrow}} % up-down ketbra
        \newcommand*{\downdown}{\ketbra{\downarrow}{\downarrow}} % down-down ketbra
\definecolor{quantum_walk_green}{RGB}{30,200,30}
\definecolor{quantum_walk_pink}{RGB}{230,0,230}
\definecolor{classical_walk_blue}{RGB}{30,200,200}
\definecolor{classical_walk_red}{RGB}{230,0,0}
\definecolor{symmetry_walk_blue}{RGB}{200,200,30}
\definecolor{symmetry_walk_yellow}{RGB}{0,0,230}



\usepackage[colorlinks]{hyperref}
\hypersetup{
    breaklinks  = true,
    colorlinks  = true,
    citecolor   = PineGreen,
    linkcolor   = MidnightBlue,
    urlcolor    = PineGreen
}

\declaretheorem[style=plain]{theorem}

\declaretheorem[style=plain,sibling=theorem]{lemma}

\AtBeginDocument{\renewcommand{\Re}{\operatorname{Re}}}
\AtBeginDocument{}

\begin{document}

\title{Classification of coined quantum walks on the line and comparison to correlated classical random walks}

\date{\today}

\author{Lukas \surname{Hantzko}}
\email{lukas.hantzko@itp.uni-hannover.de}
\affiliation{Institut für Theoretische Physik, Leibniz Universität Hannover}

\author{Lennart \surname{Binkowski}}
\email{lennart.binkowski@itp.uni-hannover.de}
\affiliation{Institut für Theoretische Physik, Leibniz Universität Hannover}

\begin{abstract}
We present a comprehensive classification of one-dimensional coined quantum walks on the infinite line, focusing on the spatial probability distributions they induce.
Building on prior results, we identify all initial coin states that lead to symmetric quantum walks for arbitrary coins, and provide a bijective parametrisation of all symmetric quantum walks modulo distributional equivalence.
Extending beyond the symmetric case, we also give a surjective parametrisation of all coined quantum walks under the same equivalence relation and a bijective parametrisation modulo equivalence of the walks' limiting distributions.

Furthermore, we derive corrected closed-form expressions for the walk amplitudes, resolving inaccuracies in previous literature, and generalise the approach to the correlated classical random walk.
This unified framework enables a direct comparison between quantum and classical dynamics.
Additionally, we discuss the asymptotic scaling of variances for both models, identifying quadratic spreading as a hallmark of non-trivial quantum walks and contrasting it with the linear behaviour of classical walks, except at the extremal points of maximal correlation.
Finally, we compare the limiting distributions arising from quantum walks with the ones in the classical case.

\end{abstract}

\maketitle

\section{\label{section:Introduction}Introduction}

Quantum walks (QWs) \cite{Kempe2003QuantumRandomWalksAnIntroductoryOverview} provide a fundamental framework for modelling coherent dynamics on discrete structures, such as graphs, and serve as an essential tool in quantum computation \cite{Schoening1999AProbabilisticAlgorithmForKSATAndConstraintSatisfactionProblems,Shenvi2003QuantumRandomWalkSearchAlgorithm,Ambainis2003QuantumWalksAndTheirAlgorithmicApplications,Magniez2007SearchViaQuantumWalk}, condensed matter \cite{Kitagawa2010ExploringTopologicalPhasesWithQuantumWalks,Asboth2014ChiralSymmetryAndBulkBoundaryCorrespondenceInPeriodicallyDrivenOneDimensionalSystems,Cedzich2018TheTopologicalClassificationOfOneDimensionalSymmetricQuantumWalks}, transport \cite{Stefanak2008RecurrencePropertiesOfUnbiasedCoinedQuantumWalksOnInfiniteDimensionalLattices,Mohseni2008EnvironmentAssistedQuantumWalksInPhotosyntheticEnergyTransfer}, statistical physics \cite{Oliveira2006DecoherenceInTwoDimensionalQuantumWalks,Chandrashekar2010DiscreteTimeQuantumWalkDynamicsAndApplications}.
Two widely studied formulations of QWs are continuous-time QWs (CTQWs), introduced by \citet{Farhi1998QuantumComputationAndDecisionTrees}, and discrete-time coined QWs (DTQWs), initiated by \citet{Aharonov1993QuantumRandomWalks} and further generalised in \cite{Aharonov2001QuantumWalksOnGraphs}.

In one dimension, DTQWs offer a tractable yet nontrivial model for investigating the interplay between interference and coherent evolution.
In this setting, a quantum particle evolves on an infinite line in discrete time steps, with its position updated conditionally based on an internal spin-degree of freedom which is subject to a unitary transformation, a unitary ``quantum coin toss'', before each step.
After each coin operation, the particle shifts conditionally: if its internal state is ‘spin-up’, it moves one step right; if ‘spin-down’, one step left.
However, it does so coherently, that is without intermediate measurements.
This setup can lead to probability distributions that exhibit characteristic features such as ballistic spreading (superlinear increase of variance) and non-Gaussian profiles.

The Hadamard walk -- where the coin operation is given by the Hadamard gate -- is the earliest example of a DTQW that has been extensively studied.
Using tools such as a path integral approach and Fourier analysis, \citet{Ambainis2001OneDimensionalQuantumWalks} were able to derive explicit formulas for the quantum state after each step of the quantum walk and with it several striking results:
Asymptotically, the quantum state's amplitudes lead to a uniform spread of probability around the origin within an interval growing proportional to the square root of the number of steps.
This further implies that the distribution's variance increases quadratically in the number of steps.

The Hadamard walk received significant attention due to its stark contrast with the symmetric classical random walk in one dimension, a well-studied textbook example of classical random walks (see, e.g., \cite{Feller1991AnIntroductionToProbabilityTheoryAndItsApplications}).
Most notably, the classical counterpart yields binomial distributions after each step such that the limiting distribution, by the de Moivre-Laplace theorem, is a Gaussian.
Especially, the probabilities are not at all uniformly distributed around the origin, but concentrate exactly in the middle.
The distributions admit a diffusive behaviour, i.e., the variance scales only linearly with the number of steps.

The early results on the Hadamard walk were soon generalised to arbitrary quantum coins and initial states, most notably by \citet{Konno2002QuantumRandomWalksInOneDimension} who determined the limiting distribution in the general case, necessary and sufficient conditions for the symmetry of the induced probability distributions for finitely many steps, and closed-form expressions for all moments of the general quantum walk.
Qualitatively, all one-dimensional coined quantum walks, except two trivial edge cases, behave similarly.
These edge cases are given by quantum coins which essentially act either as the identity or the Pauli-X gate.
For example, all non-trivial quantum coins induce probability distributions whose variance scales, at least asymptotically, quadratically with the number of steps.

While the Hadamard walk may be interpreted as the quantum analogue of the unbiased classical random walk, the analogy somewhat breaks down when considering more general quantum coins.
Biasing the classical random walk in any direction leads to asymmetry, whereas a continuum of symmetric QWs exists.
Therefore, in order to maintain comparability to the general QW, we need to enhance the classical setup.
The most natural candidate is to introduce correlations between subsequent steps of the walk and to do so symmetrically.
If the classical particle has previously made a step to the left (right), it continues to do so in the next step with some probability $p$, but flips its direction with probability $1 - p$.
Such correlated classical random walks have already been studied extensively, e.g., by \citet{Goldstein1951OnDiffusionByDiscontinuousMovementsAndOnTheTelegraphEquation} and \citet{Gillis1955CorrelatedRandomWalk}.
Their derived formulas for the probability distributions after each step as well as for the variance allow for a meaningful comparison to the quantum case.

In this article, we primarily build on Konno's work and prove several enhancements over the known literature.
Namely, we determine, for an arbitrary quantum coin, all initial coin states which render the resulting QW symmetric.
We further give a bijective parametrisation of all symmetric QWs modulo the equivalence relation of producing the same probability distributions after each step.
Under the same equivalence relation, we also provide a surjective parametrisation of all (not only symmetric) QWs.
We further provide a bijective parametrisation of all QWs modulo identical limiting distributions.

Furthermore, we reiterate recent work by \citet{Jayakody2023ClosedFormExpressionsForTheProbabilityDistributionOfQuantumWalkOnALine} which is aiming at providing closed-form expressions for the quantum state after each step of a general coined QW.
We correct a minor error in their calculations, thus producing the correct result, and extend their methodology to the correlated classical random walk in order to determine a closed-form expression for the classical probability distribution after each step.

Additionally, we discuss the most striking difference between quantum walks and correlated classical random walks in detail: the asymptotic scaling of their variances.
We thoroughly cover the edge cases in both regimes and also analyse the variances graphically.
Moreover, we investigate the limiting distributions in both frameworks for further comparisons.
Here, we also provide an example showing that there are more distributional equivalence classes of QWs than limiting distributions.
This builds directly on the previously derived closed-form expressions for the QW's states after each step.

\section{\label{section:Preliminaries}Preliminaries}

We consider the Hilbert space $\hil = \ell^{2}(\Z) \otimes \C^{2}$ of a spin-$\tfrac{1}{2}$ particle on the discrete infinite line.
We refer to $\ell^{2}(\Z)$ as the \emph{line space} and to its canonical orthonormal basis $\{\ket{j} \defcolon j \in \Z\}$ as \emph{position basis}.
$\C^{2}$ plays the role of the \emph{coin space} with orthonormal \emph{spin-z basis} $\{\up, \down\}$.
The probability of measuring the particle at the $j$-th position, $j \in \Z$, when being in the state $\ket{\psi} \in \hil$ is given by
\begin{align*}
    p(j;\, \ket{\psi}) = \abs{\braket{\psi | (\ketbra{j}{j} \otimes \one) | \psi}}^{2}.
\end{align*}

A unitary operator $C \in \unitary(\C^{2}) \subset \lo(\C^{2})$ acting on the coin space is called a (\emph{quantum}) \emph{coin}.
As usual, the conditional translation operator $T \in \unitary(\hil)$ is defined as
\begin{align}\label{equation:TranslationOperator}
    T \coloneqq \sum_{j \scriptin \Z} \ketbra{j + 1}{j} \otimes \upup + \sum_{j \scriptin \Z} \ketbra{j - 1}{j} \otimes \downdown.
\end{align}

Combining a quantum coin toss -- i.e.\ the application of some $C \in \unitary(\C^{2})$ -- with the conditional translation operator constitutes one step with the corresponding coined quantum walk.
We define the (unitary) \emph{walk operator} $W(C) \in \unitary(\hil)$ accordingly as $W(C) \coloneqq T (\one \otimes C)$.
Throughout this article, we will only consider coined quantum walks that are initialized in a product state of the form $\ket{\psi_{0}} \coloneqq \ket{0} \otimes \ket{\gamma}$, where $\ket{\gamma} \in \C^{2}$ is an arbitrary initial coin state.
That is, the particle starts in the ``middle'' of the infinite line, but with arbitrarily superposed spin states.
In this setting, a coined quantum walk is therefore uniquely determined by the \emph{coin setup} $(C, \ket{\gamma}) \in \unitary(\C^{2}) \times \C^{2}$ so that the walk's properties can be fully reduced to properties of the coin setup.
One of these properties we define in the following.

A quantum walk with setup $\mathcal{C} \coloneqq (C, \ket{\gamma}) \in \unitary(\C^{2}) \times \C^{2}$ is called (\emph{spatially}) \emph{symmetric in distribution} if, for every $j, n \in \N$, it holds that 
\begin{align*}
    p_{\mathcal{C}}(j, n) \coloneqq p\big(j;\, W(C)^{n} \ket{0} \otimes \ket{\gamma}\big) = p_{\mathcal{C}}(-j, n).
\end{align*}
That is, the symmetry condition is only enforced on the respective probability distribution $p_{\mathcal{C}}(\defdot, n): \Z \rightarrow [0, 1]$ induced by the state obtained after $n$ steps, not on the state itself.

We are further interested which coin setups give rise to the same family of induced spatial probability distributions.
We define a relation (which readily is an equivalence relation) on $\unitary(\C^{2}) \times \C^{2}$ via
\begin{align*}
    \mathcal{C}_{1} \sim_{\text{d}} \mathcal{C}_{2} :\hspace*{-3pt}\iff p_{\mathcal{C}_{1}}(j, n) = p_{\mathcal{C}_{2}}(j, n)\ \ \forall\, j \in \Z\, \forall\, n \in \N,
\end{align*}
and call such two coin setups \emph{distributionally equivalent}.

For the initial coin state $\ket{\gamma}$, we introduce the notation
\begin{align}\label{equation:CoinStateParametrisation}
    \ket{\gamma} \coloneqq \alpha \up + \beta \down \coloneqq \cos(\varphi) \up + \sin(\varphi) e^{i \xi} \down
\end{align}
with $\varphi \in [0, \pi / 2]$ and $\xi \in [0, 2 \pi)$, which parametrises all possible initial states up to global phases.

Similarly, we choose a parametrisation of all quantum coins up to global phases, thereby restricting to special unitaries, i.e.\ unitaries of determinant one:
\begin{align}\label{equation:HopfCoordinates}
\begin{split}
    C &\coloneqq a \upup + b \updown + c \downup + d \downdown \\
    &\coloneqq \cos(\theta) e^{i \frac{\phi_{1} + \phi_{2}}{2}}\hspace*{-1pt} \upup + \sin(\theta) e^{i \frac{\phi_{1} - \phi_{2}}{2}}\hspace*{-1pt} \updown \\
    &\hphantom{\coloneqq}\, - \sin(\theta) e^{-i \frac{\phi_{1} - \phi_{2}}{2}}\hspace*{-1pt} \downup + \cos(\theta) e^{-i \frac{\phi_{1} + \phi_{2}}{2}}\hspace*{-1pt} \downdown
\end{split}
\end{align}
with $\theta \in [0, 2 \pi)$ and $\phi_{1}, \phi_{2} \in [0, \pi)$.
These are also referred to as \emph{Hopf coordinates}.
Another parametrisation used, e.g., in \cite{Jayakody2023ClosedFormExpressionsForTheProbabilityDistributionOfQuantumWalkOnALine} is related to our choice by multiplying each entry with $e^{-i \frac{\phi_{1} + \phi_{2}}{2}}$ and the transformation $(\phi_{1}, \phi_{2}) \mapsto (-\phi_{2}, \pi - \phi_{1})$.
While this produces shorter coefficients, it parametrises a different, less tangible subset of all unitaries.
We call a quantum coin $C$ \emph{trivial} iff $a b c d = 0$, and denote by $\mathcal{T}(\C^{2})$ the set of all trivial coins.

According to \cite[Theorem 1]{Konno2005ANewTypeOfLimitTheoremsForTheOneDimensionalQuantumRandomWalk}, the sequence of probability distributions $p_{\mathcal{C}}(\defdot, n) / n$ has a well-defined limiting distribution with density $f_{\mathcal{C}}$ whenever the quantum coin is non-trivial.
We define a second equivalence relation on $\mathcal{U}(\C^{2}) \times \C^{2}$ via
\begin{align*}
    \mathcal{C}_{1} \sim_{\text{l}} \mathcal{C}_{2} : \iff f_{\mathcal{C}_{1}} = f_{\mathcal{C}_{2}},
\end{align*}
and call such two coin setups \emph{asymptotically distributionally equivalent}.

We denote by $\alpha_{j}$ and $\beta_{j}$ the amplitudes of $\ket{\psi_{n}}$, the state resulting from the $n$-fold application of the walk operator to the initial state $\ket{\psi_{0}}$:
\begin{align}\label{equation:WalkStateAmplitudes}
    \ket{\psi_{n}} \coloneqq \sum_{j \scriptin \Z} \big(\alpha_{j}(n) \up t+ \beta_{j}(n) \down\big) \otimes \ket{j}.
\end{align}

Additionally to quantum walks we also consider correlated classical random walks on a discrete infinite line.
The spatial information is described by probability distributions over $\Z$, i.e.\ normalised and positive elements of $\ell^{1}(\Z)$.
We assume a centralised initial distribution $p_{0} = (\ldots 0\, 0\, 1\, 0\, 0 \ldots)$ with a zeroth entry of $1$, and zeros everywhere else.
Similar to the corresponding quantum walk, the spatial distribution's spread should depend on the most recent directional information.
In order to cast this situation into the well established framework of discrete Markov chains, consider probability distributions over $\Z \times \{\uparrow, \downarrow\}$.
In analogy to the quantum case, the initial distribution should be an arbitrary distribution $r_{0}$ with $r_{0}(j, \circ) = p_{0}(j) q_{0}(\circ)$ for all $j \in \Z$ and $\circ \in \{\uparrow, \downarrow\}$ with $q_{0}$ being an arbitrary probability distribution over $\{\uparrow, \downarrow\}$.

The quantum coin is replaced by an autocorrelation of the directional information/velocity.
Given a correlation coefficient $\delta \in [-1, 1]$, the direction is flipped with probability $(1 - \delta) / 2$.
The corresponding transition matrix for the velocity part reads
\begin{align}\label{equation:VelocityTransitionMatrix}
     M  = \left(\begin{array}{c c} \frac{1 + \delta}{2} & \frac{1 - \delta}{2} \\ \frac{1 - \delta}{2} & \frac{1 + \delta}{2} \end{array}\right).
\end{align}
After adjusting the directional information, a translation analogous to \eqref{equation:TranslationOperator} adjusts the spatial information.
That is, the classical walk operator has the same structure as its quantum counterpart, but the unitary quantum coin $C$ is replaced by a doubly stochastic transition matrix $M$.

\section{\label{section:Results}Results}

Our main results concern spatially symmetric quantum walks and are extensions of Konno's work~\cite{Konno2005ANewTypeOfLimitTheoremsForTheOneDimensionalQuantumRandomWalk}.
First, we establish that an arbitrary quantum coin can be used to generate a symmetric quantum walk.

\begin{theorem}[Identification of symmetric walks]\label{theorem:IdentificationSymmetricWalks}
    For every non-trivial coin $C \in \unitary(\C^{2})$, there exist (up to a global phase) exactly two coin states $\ket{\gamma_{1}}, \ket{\gamma_{2}} \in \C^{2}$ such that $\mathcal{C}_{1} \coloneqq (C, \ket{\gamma_{1}})$ and $\mathcal{C}_{2} \coloneqq (C, \ket{\gamma_{2}})$ are coin setups for quantum walks symmetric in distribution.
    For every trivial coin there exist uncountably many coin states such that the resulting coin setups induce quantum walks symmetric in distribution.
\end{theorem}

Second, we give a bijective parametrisation of all distributional equivalence classes of symmetric quantum walks.

\begin{theorem}[Classification of symmetric walks]\label{theorem:ClassificationSymmetricWalks}
    The parametrisation
    \begin{align}\label{equation:SymmetricClassificationParametrisation}
    \begin{split}
        \theta\hspace*{-1pt} \mapsto\hspace*{-1pt} \bigg(&\hspace*{-3pt}\cos(\theta) \big(\upup\hspace*{-2pt} +\hspace*{-2pt} \downdown\big)\hspace*{-2pt} +\hspace*{-2pt} \sin(\theta) \big(\updown\hspace*{-2pt} -\hspace*{-2pt} \downup\big), \\
        &\frac{1}{\sqrt{2}} (\up + i \down)\bigg)
    \end{split}
    \end{align}
    establishes a bijection between $[0, \pi / 2]$ and the (representatives of distinct) distributional equivalence classes of symmetric coin setups.
\end{theorem}

Third, we give a (non-bijective) parametrisation of all distributional equivalence classes of quantum walks, that is also of non-symmetric ones.
In \autoref{section:LimitingDistributions} we reason why it seems substantially more difficult to derive a bijective parametrisation for all quantum walks than for only the symmetric ones.

\begin{theorem}[Classification of all walks]\label{theorem:Classification}
    The parametrisation
    \begin{align}\label{equation:ClassificationParametrisation}
        (\varphi, \xi, \theta)\hspace*{-1pt} \mapsto\hspace*{-1pt} \bigg(&\hspace*{-3pt}\cos(\theta) \big(\upup\hspace*{-2pt} +\hspace*{-2pt} \downdown\big)\hspace*{-2pt} +\hspace*{-2pt} \sin(\theta) \big(\updown\hspace*{-2pt} -\hspace*{-2pt} \downup\big), \nonumber \\
        & \cos(\varphi) \up + \sin(\varphi) e^{i \xi} \down\bigg)
    \end{align}
    surjectively maps $[0, \pi / 2] \times [0, \pi] \times [0, \pi / 2]$ to the (representatives of distinct) distributional equivalence classes of all coin setups.
\end{theorem}

Fourth, we construct a bijective parametrisation of all asymptotically distributional equivalence classes of quantum walks, again not only restricted to symmetric ones.
In \autoref{section:LimitingDistributions}, we compare this parametrisation to the one given in \autoref{theorem:Classification} and discuss differences.

\begin{theorem}[Classification of limiting distributions]\label{theorem:ClassificationLimitingDistribution}
    The parametrisation
    \begin{align}\label{equation:ClassificationLimitingDistributionParametrisation}
        (\varphi, \xi, \theta)\hspace*{-1pt} \mapsto\hspace*{-1pt} \bigg(&\hspace*{-3pt}\cos(\theta) \big(\upup\hspace*{-2pt} +\hspace*{-2pt} \downdown\big)\hspace*{-2pt} +\hspace*{-2pt} \sin(\theta) \big(\updown\hspace*{-2pt} -\hspace*{-2pt} \downup\big), \nonumber \\
        & \cos(\varphi) \up + \sin(\varphi) e^{i \xi} \down\bigg)
    \end{align}
    bijectively maps the set
    \begin{align*}
        &\{(\varphi, 0, \theta) \defcolon \theta \in (0, \pi / 2), \varphi \in [\theta / 2, \pi / 4 + \theta / 2]\} \\
        &\cup\, \{(\varphi, \pi, \theta) \defcolon \theta \in (0, \pi / 2), \varphi \in (\pi / 4 - \theta / 2, (\pi - \theta) / 2]\}
    \end{align*}
    to the (representatives of distinct) asymptotic distributional equivalence classes of all coin setups with non-trivial coins.
\end{theorem}

Furthermore, we derive a closed-form expression for the amplitudes \eqref{equation:WalkStateAmplitudes} of the quantum state after arbitrary many steps of a quantum walk.
This has been recently addressed by \citet{Jayakody2023ClosedFormExpressionsForTheProbabilityDistributionOfQuantumWalkOnALine} for more general initial states than we consider in this work.
Unfortunately, there has been a minor mistake in their calculations, rendering the final result invalid.
We give here a corrected version, following essentially the same steps.

\begin{lemma}\label{lemma:QuantumWalkClosedForm}
    Given a quantum walk with coin setup $(C, \ket{\gamma})$, where $C$ is a quantum coin $C$ with \hyperref[equation:HopfCoordinates]{Hopf coordinates} $\theta \in [0, 2 \pi)$ and $\phi_{1}, \phi_{2} \in [0, \pi)$, and $\ket{\gamma} = \alpha \up + \beta \down$ is a coin state, the amplitudes of the state \eqref{equation:WalkStateAmplitudes} after $n \in \N$ applications of the walk operator read
    \begin{align*}
    \begin{split}
        \alpha_{j}(n) & = e^{i (j - 1) \frac{\phi_{1} + \phi_{2}}{2}} \Bigg(\sum_{h = 0}^{n} \kappa_{n, j}^{h} \cos^{h}(\theta) e^{i \frac{\phi_{1} + \phi_{2}}{2}} (-1)^{\frac{n - h}{2}} \alpha \\
        & + \sum_{h = 0}^{n - 1} \kappa_{n - 1, j + 1}^{h} \cos^{h + 1}(\theta) e^{i \frac{\phi_{1} + \phi_{2}}{2}} (-1)^{\frac{n + 1 - h}{2}} \alpha \\
        & + \sum_{h = 0}^{n - 1} \kappa_{n - 1, j - 1}^{h} \cos^{h}(\theta) \sin(\theta) e^{i \frac{\phi_{1} - \phi_{2}}{2}} (-1)^{\frac{n - 1 - h}{2}} \beta\Bigg), \\
        \beta_{j}(t) & = e^{i (j + 1) \frac{\phi_{1} + \phi_{2}}{2}} \Bigg(\sum_{h = 0}^{n} \kappa_{n, j}^{h} \cos^{h}(\theta) e^{-i \frac{\phi_{1} + \phi_{2}}{2}}\hspace*{-1pt} (-1)^{\frac{n - h}{2}} \beta \\
        & + \sum_{h = 0}^{n - 1} \kappa_{n - 1, j + 1}^{h} \cos^{h}(\theta) \sin(\theta) e^{-i \frac{\phi_{1} - \phi_{2}}{2}} (-1)^{\frac{n - 1 - h}{2}} \alpha \\
        & + \sum_{h = 0}^{n - 1} \kappa_{n - 1, j - 1}^{h} \cos^{h + 1}(\theta) e^{-i \frac{\phi_{1} + \phi_{2}}{2}} (-1)^{\frac{n + 1 - h}{2}} \beta\Bigg)
    \end{split}
    \end{align*}
    with combinatorial prefactors
    \begin{align}\label{equation:CombinatorialPrefactor}
        \kappa_{n, j}^{h} = \frac{\frac{n + h}{2}!}{\frac{n - h}{2}! \frac{h - j}{2}! \frac{h + j}{2}!}.
    \end{align}
\end{lemma}

In analogy, we also derive a closed-form expression for the components of the joint spatial-directional probability distribution of the correlated random walk.
Due to the similar setup, we can use the exact same tools as for the quantum case.
This complements a result by \citet{Gillis1955CorrelatedRandomWalk} who derived an alternative, more compact expression.

\begin{lemma}\label{lemma:ClassicalWalkClosedForm}
    Let $q_{0} = (\alpha, \beta)$ be an arbitrary probability distribution over $\{\uparrow, \downarrow\}$, i.e.\ $0 \leq \alpha, \beta \leq 1$ with $\alpha = 1 - \beta$.
    Let $\delta \in [-1, 1]$ parametrise the velocity transition matrix \eqref{equation:VelocityTransitionMatrix} and let $r_{n}$ be the probability distribution over $\Z \times \{\uparrow, \downarrow\}$ after $n \in \N$ steps of the corresponding correlated random walk with components $r_{n}(j, \uparrow) \coloneqq \alpha_{j}(n)$ and $r_{n}(j, \downarrow) \coloneqq \beta_{j}(n)$, $j \in \Z$.
    Then these components are given by
    \begin{align*}
    \begin{split}
        \alpha_{j}(n) & = \sum_{h = 0}^{n} \kappa_{n, j}^{h} \left(\frac{1 + \delta}{2}\right)^{h} (-\delta)^{\frac{n - h}{2}} \alpha \\
        & - \sum_{h = 0}^{n - 1} \kappa_{n - 1, j + 1}^{h} \left(\frac{1 + \delta}{2}\right)^{h + 1} (-\delta)^{\frac{n - 1 - h}{2}} \alpha \\
        & + \sum_{h = 0}^{n - 1} \kappa_{n - 1, j - 1}^{h} \left(\frac{1 + \delta}{2}\right)^{h} \frac{1 - \delta}{2} (-\delta)^{\frac{n - 1 - h}{2}} \beta, \\
        \beta_{j}(n) &= \sum_{h = 0}^{n} \kappa_{n, j}^{h} \left(\frac{1 + \delta}{2}\right)^{h} (-\delta)^{\frac{n - h}{2}} \alpha \\
        & - \sum_{h = 0}^{n - 1} \kappa_{n - 1, j - 1}^{h} \left(\frac{1 + \delta}{2}\right)^{h + 1} (-\delta)^{\frac{n - 1 - h}{2}} \beta \\
        & + \sum_{h = 0}^{n - 1} \kappa_{n - 1, j + 1}^{h} \left(\frac{1 + \delta}{2}\right)^{h} \frac{1 - \delta}{2} (-\delta)^{\frac{n - 1 - h}{2}} \alpha
    \end{split}
    \end{align*}
    with the same combinatorial prefactors $\kappa_{n, j}^{h}$ as in \eqref{equation:CombinatorialPrefactor}.
\end{lemma}

\section{\label{section:Methods}Methods}

\subsection{\label{subsection:Classification}Classification}

Our main three theorems can be compactly proven with the aid of Konno's previous work \cite{Konno2005ANewTypeOfLimitTheoremsForTheOneDimensionalQuantumRandomWalk}.
Therein, Konno derives closed-form expressions for the spatial probability distributions after each application of the walk operator, utilising the path integral approach pioneered in \cite{Ambainis2001OneDimensionalQuantumWalks} for the Hadamard walk.
Konno uses those expressions to determine the characteristic function of each induced probability distribution.
Most importantly, a tangible expression for the expectation value yields the necessary and sufficient conditions for spatial symmetry.

\begin{proof}[Proof of \autoref{theorem:IdentificationSymmetricWalks}]
    This result is an easy consequence of \cite[Theorem 6]{Konno2005ANewTypeOfLimitTheoremsForTheOneDimensionalQuantumRandomWalk} which establishes:
    Given a non-trivial quantum coin $C$ (i.e.\ $a b c d \neq 0$) and a coin state $\ket{\gamma} = \alpha \up + \beta \down$, the quantum walk with coin setup $(C, \ket{\gamma})$ is symmetric in distribution iff $\abs{\alpha} = \abs{\beta} = 1 / \sqrt{2}$ and $a \alpha \overline{b \beta} + \overline{a \alpha} b \beta = 0$.
    Using our coordinates for the coin state the first condition enforces $\alpha = 1 / \sqrt{2}$ and $\beta = e^{i \xi} / \sqrt{2}$ for some $\xi \in [0, 2 \pi)$.
    Inserting this and the \hyperref[equation:HopfCoordinates]{Hopf coordinates} into the second condition yields
    \begin{align*}
        a \overline{b} e^{-i \xi} &= - \overline{a} b e^{i \xi} \iff e^{i (\phi_{1} + \phi_{2})} = - e^{i (\phi_{1} - \phi_{2})} e^{2i \xi} \\
        &\iff \phi_{1} + \phi_{2} \equiv \pi + \phi_{1} - \phi_{2} + 2 \xi \mod 2 \pi \\
        &\iff \phi_{2} \equiv \tfrac{\pi}{2} + \xi \mod \pi
    \end{align*}
    which, for given $\phi_{2} \in [0, \pi)$, is solved by exactly two possible values for $\xi \in [0, 2\pi)$.

    Also trivial coins can be used to generate symmetric quantum walks.
    Since $\abs{a} = \abs{d}$ and $\abs{b} = \abs{c}$, but $\det(C) = a d - b c \neq 0$, the assumption $a b c d = 0$ yields two distinct cases:
    either $a = d = 0$ or $b = c = 0$.
    In both cases, the second condition for symmetry, namely $a \alpha \overline{b \beta} + \overline{a \alpha} b \beta = 0$, is automatically fulfilled which lifts the restriction on the $\xi$-parameter of the coin state.
    The restriction $\abs{\alpha} = \abs{\beta} = 1 / \sqrt{2}$ remains.
    Therefore, a trivial coin $C$ can be paired up with any coin state $\ket{\gamma} = (\up + e^{i \xi} \down) / \sqrt{2}$, $\xi \in [0, 2 \pi)$ to generate a spatially symmetric quantum walk.
\end{proof}

\begin{proof}[Proof of \autoref{theorem:ClassificationSymmetricWalks}]
    First of all, since global phases in the coin state and the coin do not affect the spatial probability distribution, the joint parametrisation of coin state \eqref{equation:CoinStateParametrisation} and \eqref{equation:HopfCoordinates} coin reaches all possible coined quantum walks (not only spatially symmetric ones).
    In \cite[Lemma 3]{Konno2005ANewTypeOfLimitTheoremsForTheOneDimensionalQuantumRandomWalk}, Konno gives explicit expressions for the spatial distribution after each step of the quantum walk.
    Apart from the position $j \in \Z$ and the number of steps $n \in \N$, this distribution only depends on $\abs{\alpha}^{2} = \cos^{2}(\varphi)$, $\abs{\beta}^{2} = \sin^{2}(\varphi)$, $\abs{a}^{2} = \cos^{2}(\theta)$, $\abs{b}^{2} = \sin^{2}(\theta)$, as well as
    \begin{align*}
        a \alpha \overline{b \beta} + \overline{a \alpha} b \beta = 2 \cos(\theta) \sin(\theta) \cos(\varphi) \sin(\varphi) \cos(\xi - \phi_{2}).
    \end{align*}
    That is, the Hopf coordinate $\phi_{1}$ does not impact the distribution at all.
    Phrased differently, for every coin setup $\mathcal{C} = (C, \ket{\gamma})$, there is a second coin setup $\tilde{\mathcal{C}} =(\tilde{C}, \ket{\gamma})$ with $\phi_{1} = 0$ such that $\mathcal{C} \sim_{\text{d}} \tilde{\mathcal{C}}$.
    
    Since we specifically consider symmetric quantum walks, $\varphi \in [0, \pi / 2]$ is forced to have the value $\pi / 4$ in order to ensure the necessary condition $\abs{\alpha}^{2} = \abs{\beta}^{2} = 1 / 2$, hence it is also eliminated as a free parameter.
    The imposed spatial symmetry of the walk further fixes $a \alpha \overline{b \beta} + \overline{a \alpha} b \beta = 0$ such that $\theta$ is the only remaining Hopf coordinate which impacts the spatial probability distribution (via $\abs{a}^{2}$ and $\abs{b}^{2}$).
    Since $\cos^{2}$ and $\sin^{2}$ are $\pi$-periodic, we can first restrict $\theta$ to the interval $[0, \pi)$.
    Since $(\cos^{2}(\pi / 2 - \theta), \sin^{2}(\pi / 2 - \theta)) = (\cos^{2}(\pi / 2 + \theta), \sin^{2}(\pi / 2 + \theta))$, it even suffices to restrict $\theta$ to $[0, \pi / 2]$ with the overall restriction/identification function
    \begin{align*}
    \begin{split}
        \varrho : [0, 2 \pi) &\rightarrow [0, \pi / 2] \\
        x &\mapsto \begin{cases}
            x, & \text{if } 0 \leq x < \pi / 2, \\
            \pi - x, & \text{if } \pi / 2 \leq x < \pi, \\
            x - \pi, & \text{if } \pi \leq x < 3 \pi / 2, \\
            2 \pi - x, & \text{otherwise}.
        \end{cases}
    \end{split}
    \end{align*}
    Furthermore, due to \autoref{theorem:IdentificationSymmetricWalks}, $\xi - \phi_{2}$ is equal to $\pi (1 \pm 1 / 2)$ which is, e.g., fulfilled for $\xi = \pi / 2$ and $\phi_{2} = 0$.

    In summary, for every coin setup $\mathcal{C} = (C, \ket{\gamma})$ with $\theta = \theta_{C}$ inducing a spatially symmetric quantum walk, there exists a second coin setup $\tilde{\mathcal{C}} = (\tilde{C}, \ket{\tilde{\gamma}})$ such that $\mathcal{C} \sim_{\text{d}} \tilde{\mathcal{C}}$, $\ket{\tilde{\gamma}}$ has \hyperref[equation:CoinStateParametrisation]{parameters} $(\varphi, \xi) = (\pi / 4, \pi / 2)$, and $\tilde{\mathcal{C}}$ has \hyperref[equation:HopfCoordinates]{Hopf coordinates} $(\theta, \phi_{1}, \phi_{2}) = (\varrho(\theta_{C}), 0, 0)$, establishing surjectivity of the parametrisation \eqref{equation:SymmetricClassificationParametrisation}.
    Injectivity readily follows from the expressions for the boundary probabilities $p_{\mathcal{C}}(n, n)$, $n \in \N$.
    For symmetric walks those read, according to \cite[Lemma 3]{Konno2005ANewTypeOfLimitTheoremsForTheOneDimensionalQuantumRandomWalk},
    \begin{align*}
        p_{\mathcal{C}}(n, n) &= \abs{a}^{2 (n - 1)} (\abs{b}^{2} + \abs{a}^{2}) / 2 \\
        &= \big(\hspace*{-1pt}\cos^{2 n}(\theta) + \cos^{2 (n - 1)}(\theta) \sin^{2}(\theta)\big) / 2 \\
        & = \big(\hspace*{-1pt}\cos^{2 n}(\theta) + \cos^{2 (n - 1)}(\theta) (1 - \cos^{2}(\theta))\big) / 2 \\
        &= \cos^{2}(\theta) / 2.
    \end{align*}
    Since $\cos^{2}$ is injective on the interval $[0, \pi / 2]$, this already shows injectivity of the parametrisation \eqref{equation:SymmetricClassificationParametrisation}.
\end{proof}

\begin{proof}[Proof of \autoref{theorem:Classification}]
    Recall that, according to \cite[Lemma 3]{Konno2005ANewTypeOfLimitTheoremsForTheOneDimensionalQuantumRandomWalk}, the spatial probability distributions only depend on $\cos^{2}(\varphi)$, $\sin^{2}(\varphi)$, $\cos^{2}(\theta)$, $\sin^{2}(\theta)$, and $\kappa \coloneqq \cos(\theta) \sin(\theta) \cos(\varphi) \sin(\varphi) \cos(\xi - \phi_{2})$.
    Especially, $\phi_{1}$ can again be eliminated as a free parameter and may be set to $0$.
    In order to reach all possible values for the first four terms, it suffices to restrict again $\theta$ to $[0, \pi / 2]$ (the domain of $\varphi$ is already given by $[0, \pi / 2]$ due to the irrelevance of global phases).
    The range of $\kappa$, even with unrestricted parameter domains, is given by $[-1 / 4, 1 / 4]$ since $\range(\cos \cdot \sin) = [-1 / 2, 1 / 2]$ and $\range(\cos) = [-1, 1]$.
    The restriction of $\theta$ to $[0, \pi / 2]$ reduces the range of $\cos \cdot \sin$ to the non-negative half, respectively.
    By fixing $\phi_{2} = 0$ and restricting $\xi$ to $[0, \pi]$, the cosine still reaches its entire range which, in turn, suffices to obtain all possible values for $\kappa$ as well.

    In summary, for every coin setup $\mathcal{C} = (C, \ket{\gamma})$ with $\varphi = \varphi_{\gamma}$, $\xi = \xi_{\gamma}$, and $\theta = \theta_{C}$, there exists a second coin setup $\tilde{\mathcal{C}} = (\tilde{C}, \ket{\tilde{\gamma}})$ such that $\mathcal{C} \sim_{\text{d}} \tilde{\mathcal{C}}$, $\ket{\tilde{\gamma}}$ has \hyperref[equation:CoinStateParametrisation]{parameters} $(\varphi, \xi) = (\varphi_{\gamma}, \xi_{\gamma})$ if $\xi_{\gamma} \leq \pi$ and $(\varphi, \xi) = (\varphi_{\gamma}, \pi - \xi_{\gamma})$ otherwise, and $\tilde{C}$ has \hyperref[equation:HopfCoordinates]{Hopf coordinates} $(\theta, \phi_{1}, \phi_{2}) = (\varrho(\theta_{C}), 0, 0)$, establishing surjectivity of the parametrisation \eqref{equation:ClassificationParametrisation}.
\end{proof}

\begin{proof}[Proof of \autoref{theorem:ClassificationLimitingDistribution}]
    One of Konno's main results is the derivation of the limiting distribution of the sequence $p_{\mathcal{C}}(\defdot, n) / n$ for all non-trivial coin setups $\mathcal{C}$, extending previous results for the Hadamard walk \cite[Theorem 2]{Ambainis2001OneDimensionalQuantumWalks}.
    Given the usual parametrisation of coin state \eqref{equation:CoinStateParametrisation} and quantum coin \eqref{equation:HopfCoordinates}, \cite[Theorem 1]{Konno2005ANewTypeOfLimitTheoremsForTheOneDimensionalQuantumRandomWalk} states that the normalised limiting distribution has the density
    \begin{align}\label{equation:LimitingDensity}
        f_{\mathcal{C}}(x) = \frac{\sqrt{1 - \abs{a}^{2}} (1 - \lambda_{\mathcal{C}} x)}{\pi (1 - x^{2}) \sqrt{\abs{a}^{2} - x^{2}}}
    \end{align}
    with
    \begin{align}\label{equation:Lambda}
        \lambda_{\mathcal{C}} = \abs{\alpha}^{2} - \vert\beta\vert^{2} + \frac{a \alpha \overline{b \beta} + \overline{a \alpha} b \beta}{\abs{a}^{2}}.
    \end{align}
    for $x \in (-\abs{a}, \abs{a})$ and $0$ elsewhere.

    First, notice that all the representatives of the distributional equivalence classes of coin setups have only real-valued coefficients.
    Using this and \hyperref[equation:HopfCoordinates]{Hopf coordinates}, we simplify \eqref{equation:Lambda} to
    \begin{align}
        \lambda_{\mathcal{C}} &= \cos^{2}(\varphi) - \sin^{2}(\varphi) + 2 \tan(\theta) \Re(\cos(\varphi) \sin(\varphi) e^{-i \xi}) \nonumber \\
        &= \cos(2 \varphi) + \sin(2 \varphi) \tan(\theta) \cos(\xi). \label{equation:LambadeSimplified}
    \end{align}
    Using the harmonic addition theorem, we obtain the alternative form
    \begin{align*}
        \lambda_{\mathcal{C}} = \sqrt{1\hspace*{-1pt} +\hspace*{-1pt} \tan^{2}\hspace*{-1pt}(\theta) \cos^{2}(\xi)} \cos(2 \varphi\hspace*{-1pt} -\hspace*{-1pt} \tan^{\hspace*{-1pt}-1}\hspace*{-1pt}(\tan(\theta) \cos(\xi))).
    \end{align*}
    
    Now observe that no two unequal pairs of $(\theta, \lambda_{\mathcal{C}})$ can yield the same density $f_{\mathcal{C}}$.
    This easily follows by expanding the expression \eqref{equation:LimitingDensity} and grouping the terms in powers of $x$.
    This already forces the domain of $\theta$ to be the full interval $(0, \pi / 2)$.
    Second, $\lambda_{\mathcal{C}}$ itself depends on $\theta$.
    In the following, we will, for fixed $\theta$, derive two pairs for the parameters $\varphi$ and $\xi$ which yield the minimal and maximal possible value for $\lambda_{\mathcal{C}}$.
    
    For the moment we also fix a value for $\xi \in [0, \pi]$ beside $\theta$ and let $\nu \coloneqq \tan(\theta) \cos(\xi)$.
    Note that since $\tan(\theta) > 0$, $\nu$ is positive for $\xi \in [0, \pi / 2)$, zero for $\xi = \pi / 2$, and negative for $\xi \in (\pi / 2, \pi]$.
    The minima and maxima of $\lambda_{\mathcal{C}}(\varphi; \nu) = \sqrt{1 + \nu^{2}} \cos(2 \varphi - \tan^{-1}(\nu))$ are readily given by $\pm \sqrt{1 + \nu^{2}}$, respectively, and attained alternately at $\varphi_{\min / \max}(\nu) = (\tan^{-1}(\nu) + k \pi) /2$, $k \in \Z$.
    However, for $\nu \neq 0$, only one of these extrema lies within the interval $[0, \pi / 2]$:
    For $\nu > 0$, $k = 0$ and $\varphi_{\max}(\nu) = \tan^{-1}(\nu) / 2$ is a maximiser, while for $\nu < 0$, $k = 1$ and $\varphi_{\min}(\nu) = (\tan^{-1}(\nu) + \pi) / 2$ is a minimiser.
    The boundary values $\varphi \in \{0, \pi / 2\}$ constitute the respective opposite extremum on $[0, \pi / 2]$:
    For $\nu > 0$, $\varphi_{\min}(\nu) = \pi / 2$ is a minimiser yielding a minimum of $-1$, while for $\nu < 0$, $\varphi_{\max}(\nu) = 0$ is a maximiser yielding a maximum of $1$.
    Lastly, for $\nu = 0$, $\varphi_{\max}(0) = 0$ maximises $\lambda_{\mathcal{C}}(\varphi; \nu)$ to be $1$ and $\varphi_{\min}(0) = \pi / 2$ minimises it with value $-1$.
    
    Now, we vary $\xi$ for fixed $\theta$.
    Clearly, $\lambda_{\max}(\nu)$ is maximised within the regime $\xi \in [0, \pi / 2)$ when $\nu$ is maximal which, in turn, is given by $\nu = \tan(\theta)$ for $\xi = 0$.
    Analogously, $\lambda_{\min}(\nu)$ is minimised within the regime $\xi \in (\pi / 2, \pi]$ when $\abs{\nu}$ is maximal which is granted by letting $\xi = \pi$.
    The respective optimisers are
    \begin{align*}
        (\varphi_{\min}, \varphi_{\max}) = \begin{cases}
            (\pi / 2, \theta / 2), & \text{if } \xi = 0, \\
            ((\pi - \theta) / 2, 0), & \text{if } \xi = \pi .
        \end{cases}
    \end{align*}
    Readily, $\lambda_{\mathcal{C}}(\defdot, 0, \theta)$ is strictly monotonically decreasing between its maximum and minimum, while $\lambda_{\mathcal{C}}(\defdot, \pi, \theta)$ is strictly monotonically increasing between its minimum and maximum.
    
    In summary, this means that restricting the parameter triple $(\varphi, \xi, \theta)$ from $(0, \pi / 2) \times [0, \pi] \times [0, \pi]$ to
    \begin{align*}
        &\{(\varphi, 0, \theta) \defcolon \theta \in (0, \pi / 2), \varphi \in [\theta / 2, \pi / 2]\} \\
        &\cup\, \{(\varphi, \pi, \theta) \defcolon \theta \in (0, \pi / 2), \varphi \in [0, (\pi - \theta) / 2]\}
    \end{align*}
    still reaches all limiting distributions since the two disjoint components cover all possible $(\theta, \lambda_{\mathcal{C}})$-pairs.
    However, for $\lambda_{\mathcal{C}} \in [-1, 1]$ there are precisely two triples, one from each component, covering the respective $(\theta, \lambda_{\mathcal{C}})$-pair.
    We can easily remove the overlap by restricting the range in $\lambda_{\mathcal{C}}$ of the first component to $[0, \sqrt{1 + \tan^{2}(\theta)}]$ and the one of the second component to $[-\sqrt{1 + \tan^{2}(\theta)}, 0)$.
    The root in $\varphi$ of $\lambda_{\mathcal{C}}(\varphi, 0, \theta)$ is given by $\varphi_{0}(0, \theta) = \pi / 4 + \theta / 2 $, while the root of $\lambda_{\mathcal{C}}(\varphi, \xi, \theta)$ is given by $\varphi_{0}(0, \theta) = \pi / 4 - \theta / 2$.
    Therefore, we have just shown that the restriction of the parameter triple $(\varphi, \xi, \theta)$ to
    \begin{align*}
        &\{(\varphi, 0, \theta) \defcolon \theta \in (0, \pi / 2), \varphi \in [\theta / 2, \pi / 4 + \theta / 2]\} \\
        &\cup\, \{(\varphi, \pi, \theta) \defcolon \theta \in (0, \pi / 2), \varphi \in (\pi / 4 - \theta / 2, (\pi - \theta) / 2]\}
    \end{align*}
    yields a bijective parametrisation of the distinct limiting distributions of all non-trivial quantum walks.
\end{proof}

\subsection{\label{subsection:ClosedFormExpressions}Closed-form expressions}

We follow the established framework of Fourier analysis for the quantum walk as an alternative to the path integral approach.
This technique has first been applied by \citet{Nayak2000QuantumWalkOnTheLine} to the Hadamard walk and is also suitable for more general quantum walks (see also \cite{Ambainis2001OneDimensionalQuantumWalks}).
The Fourier analysis allows us to retrace the $n$-fold application of the walk operator to simply applying the $n$-th power of $2 \times 2$ matrices.
In order to obtain explicit formulas for the latter, we utilise the Fibonacci-Horner method \cite{Taher2006FibonacciHornerDecompositionOfTheMatrixExponentialAndTheFundamentalSystemOfSolutions} as has been also used in \cite{Jayakody2023ClosedFormExpressionsForTheProbabilityDistributionOfQuantumWalkOnALine}.
We also extend this methodology to the correlated classical random walk which behaves mathematically similar enough to be treated with the very same techniques.

First, we define the (spatial) Fourier transform $F$ which only acts non-trivially on the spatial component of $\hil$, i.e.\ on $\ell^{2}(\Z)$.
In order to perform subsequent calculations more compactly, we also include the isomorphism $L^{2}([-\pi, \pi]) \otimes \C^{2} \cong L^{2}([-\pi, \pi], \C^{2})$ within $F$.
This allows us to separate both spin components in two vector components and to tackle both components in a single equation.
The entire Fourier transform thus reads
\begin{align*}
\begin{split}
    F : \hil &\rightarrow L^{2}([-\pi, \pi], \C^{2}) \\
    \ket{\psi} &\mapsto \tilde{\psi} \text{ with } \tilde{\psi}(k) \coloneqq \sum_{j \scriptin \Z} e^{i k j} \begin{pmatrix}
        \braket{j, \uparrow | \psi} \\
        \braket{j, \downarrow | \psi}
    \end{pmatrix}.
\end{split}
\end{align*}

Before applying $F$ to the states $\ket{\psi_{n}}$, we write out explicitly the relation between $\ket{\psi_{n}}$ and its successor state $\ket{\psi_{n + 1}} = W(C) \ket{\psi_{n}}$:
\begin{align*}
    \ket{\psi_{n + 1}}\hspace*{-1pt} =\hspace*{-1pt} \sum_{j \scriptin \Z} \ket{j}\hspace*{-1pt} \otimes\hspace*{-1pt} \big((a \braket{j\hspace*{-1pt} -\hspace*{-1pt} 1, \uparrow | \psi_{n}} + b \braket{j\hspace*{-1pt} -\hspace*{-1pt} 1, \downarrow | \psi_{n}})\hspace*{-1pt} \otimes\hspace*{-1pt} \up \\
    + (c \braket{j\hspace*{-1pt} +\hspace*{-1pt} 1, \uparrow | \psi_{n}} + d \braket{j\hspace*{-1pt} +\hspace*{-1pt} 1, \downarrow | \psi_{n}})\hspace*{-1pt} \otimes\hspace*{-1pt} \down\hspace*{-2pt}\big).
\end{align*}

This expression for $\ket{\psi_{n + 1}}$ is well suited for applying $F$ while keeping track of the dependencies on the previous state.
First, notice that the emerging double sum over spatial indices collapses to a single sum due to the orthogonality of the states $\ket{j, \circ} \coloneqq \ket{j} \otimes \ket{\circ}$, $j \in \Z$ and $\circ \in \{\uparrow, \downarrow\}$.
Second, by pulling out a prefactor of $e^{i k}$ (resp. $e^{-i k}$), the Fourier-transformed $\ket{\psi_{n + 1}}$ may be written as a linear transform (i.e.\ a $2 \times 2$ matrix) of the Fourier-transformed predecessor $\ket{\psi_{n}}$.
The entire calculation is given by
\begin{align*}
    &\tilde{\psi}_{n + 1}(k) \\
    &\ = \sum_{j \scriptin \Z} e^{i k j} \begin{pmatrix}
        a \braket{j - 1, \uparrow | \psi_{n}} + b \braket{j - 1, \downarrow | \psi_{n}} \\
        c \braket{j + 1, \uparrow | \psi_{n}} + d \braket{j + 1, \downarrow | \psi_{n}}
    \end{pmatrix} \\
    &\ = e^{i k} \sum_{j \scriptin \Z} e^{i k (j - 1)} \begin{pmatrix}
        a \braket{j - 1, \uparrow | \psi_{n}} + b \braket{j - 1, \downarrow | \psi_{n}} \\
        0
    \end{pmatrix} \\
    &\quad + e^{-i k} \sum_{j \scriptin \Z} e^{i k (j + 1)} \begin{pmatrix}
        0 \\
        c \braket{j + 1, \uparrow | \psi_{n}} + d \braket{j + 1, \downarrow | \psi_{n}}
    \end{pmatrix} \\
    &\ = e^{i k} \sum_{j \scriptin \Z} e^{i k (j - 1)} \begin{pmatrix}
        a & b \\
        0 & 0
    \end{pmatrix} \begin{pmatrix}
         \braket{j - 1, \uparrow | \psi_{n}} \\
         \braket{j - 1, \downarrow | \psi_{n}}
    \end{pmatrix} \\
    &\quad + e^{-i k} \sum_{j \scriptin \Z} e^{i k (j + 1)} \begin{pmatrix}
        0 & 0 \\
        c & d
    \end{pmatrix} \begin{pmatrix}
         \braket{j + 1, \uparrow | \psi_{n}} \\
         \braket{j + 1, \downarrow | \psi_{n}}
    \end{pmatrix} \\
    &\ = \begin{pmatrix}
        e^{i k} a & e^{i k} b \\
        e^{-i k} c & e^{-i k} d
    \end{pmatrix} \tilde{\psi}_{n}(k) \eqqcolon C_{k} \tilde{\psi}_{n}(k).
\end{align*}

By induction, this establishes $\tilde{\psi_{n}}(k) = C_{k}^{n} \tilde{\psi_{0}}(k)$.
Note that we have not used any specific property of $C$ such as unitarity.
A completely analogous Fourier analysis may be conducted for the velocity transition matrix $M$ of the correlated classical random walk, as defined in \eqref{equation:VelocityTransitionMatrix}.
Let $\tilde{r}_{n}$ denote the Fourier-transformed probability distribution after $n$ steps.
It then holds that
\begin{align}
    \tilde{r}_{n}(k)\hspace*{-2pt} =\hspace*{-2pt} M_{k}^{n} \tilde{r}_{0}(k), \text{with } M_{k}\hspace*{-2pt} \coloneqq\hspace*{-2pt} \begin{pmatrix}
        e^{i k} \tfrac{1 + \delta}{2} & e^{i k} \tfrac{1 - \delta}{2} \\
        e^{-i k} \tfrac{1 - \delta}{2} & e^{-i k} \tfrac{1 + \delta}{2}
    \end{pmatrix}\hspace*{-2pt}.
\end{align}

For both the classical and the quantum case, studying the dynamics in the Fourier space therefore reduces to calculating integer powers of $2 \times 2$ matrices.
While standard matrix diagonalisation would suffice, this task is more elegantly tackled by the Fibonacci-Horner decomposition~\cite{Taher2006FibonacciHornerDecompositionOfTheMatrixExponentialAndTheFundamentalSystemOfSolutions} which, in turn, is a consequence of the Cayley-Hamilton theorem (every square matrix is a root of its characteristic polynomial).
For a $2 \times 2$ matrix $A$ with characteristic polynomial $\chi(A) = \lambda^{2} - c_{0} \lambda - c_{1}$, the Fibonacci-Horner decomposition of the $n$-th power reads
\begin{align*}
    A^{n} = f_{n} \one + f_{n - 1} (A - c_{0} \one)
\end{align*}
with coefficients
\begin{align}
    f_{n} &= \sum_{h_{0} + 2 h_{1} = n} \frac{\left(h_{0} + h_{1}\right)!}{h_{0}! h_{1}!} c_{0}^{h_{0}} c_{1}^{h_{1}} \nonumber \\
    & = \sum_{h = 0}^{n} \frac{\left(\frac{n + h}{2}\right)!}{h! \left(\frac{n - h}{2}\right)!} c_{0}^{h} c_{1}^{\frac{n - h}{2}}. \label{equation:FibonacciCoefficient}
\end{align}

\begin{widetext}
Consider first the Fourier-transformed quantum coin $C_{k}$ in \hyperref[equation:HopfCoordinates]{Hopf coordinates}
\begin{align*}
    C_{k} = \begin{pmatrix}
        \cos(\theta) e^{i \left(k + \frac{\phi_{1} + \phi_{2}}{2}\right)} & \sin(\theta)e^{i \left(k + \frac{\phi_{1} - \phi_{2}}{2}\right)} \\
        \sin(\theta) e^{-i \left(k + \frac{\phi_{1} - \phi_{2}}{2}\right)} & \cos(\theta) e^{-i \left(k + \frac{\phi_{1} + \phi_{2}}{2}\right)}
    \end{pmatrix}
\end{align*}
with its characteristic polynomial
\begin{align*}
    \chi(\lambda) = \lambda^{2} - \cos(\theta) \Big(e^{i \left(k + \frac{\phi_{1} + \phi_{2}}{2}\right)} + e^{-i \left(k + \frac{\phi_1 + \phi_2}{2}\right)}\Big) \lambda + 1.
\end{align*}
Plugging those coefficients into \eqref{equation:FibonacciCoefficient} yields
\begin{align*}
    f_{n} = \sum_{h = 0}^{n} (-1)^{\frac{n - h}{2}} \frac{\left(\frac{n + h}{2}\right)!}{h! \left(\frac{n - h}{2}\right)!} \cos^{h}(\theta) \sum_{g = 0}^{h} \binom{h}{g} e^{i (2g - h) \left(k + \frac{\phi_{1} + \phi_{2}}{2}\right)}.
\end{align*}
Applying the Fibonacci-Horner decomposition of $C_{k}^{n}$ to $\tilde{\psi}_{0}(k) = (\alpha, \beta)^{\text{T}}$ gives
\begin{align*}
    \tilde{\psi}_{n}(k) = \begin{pmatrix}
        f_{n} - f_{n - 1} \cos(\theta) e^{-i \left(k + \frac{\phi_{1} + \phi_{2}}{2}\right)} & f_{n - 1} \sin(\theta) e^{i \left(k + \frac{\phi_{1} + \phi_{2}}{2}\right)} \\
        f_{n - 1} \sin(\theta) e^{-i \left(k + \frac{\phi_{1} + \phi_{2}}{2}\right)} & f_{n} - f_{n - 1} \cos(\theta) e^{i \left(k + \frac{\phi_{1} + \phi_{2}}{2}\right)}
    \end{pmatrix} \begin{pmatrix}
        \alpha_{\vphantom{n}} \vphantom{e^{\left(\frac{\phi_{1}}{2}\right)}} \\
        \beta_{\vphantom{n}} \vphantom{e^{\left(\frac{\phi_{1}}{2}\right)}}
    \end{pmatrix}.
\end{align*}
During this step a small error occurred in \cite{Jayakody2023ClosedFormExpressionsForTheProbabilityDistributionOfQuantumWalkOnALine}, leading to ultimately incorrect expressions for the quantum walk amplitudes.
\end{widetext}

The final step in deriving the spatial amplitudes is to apply the inverse Fourier transform to $\tilde{\psi}_{n}$ which looks like
\begin{align}\label{equation:InverseFourierTransform}
    \begin{pmatrix}
        \braket{j, \uparrow | \psi} \\
        \braket{j, \downarrow | \psi}
    \end{pmatrix} = \int_{-\pi}^{\pi} \frac{\text{d} k}{2 \pi} e^{-i k j} \tilde{\psi}_{n}(k).
\end{align}
We omit this lengthy final calculation which mostly evolves around applying the identity
\begin{align*}
    \int_{-\pi}^{\pi} \frac{\text{d}k}{2 \pi} e^{i k x} = \delta(x)
\end{align*}
multiple times.
The outcome then indeed establishes \autoref{lemma:QuantumWalkClosedForm}.

We now proceed analogously for the correlated classical random walk.
The Fourier-transformed transition matrix is given by
\begin{align}
    M_{k} = \begin{pmatrix}
        e^{i k} \frac{1 + \delta}{2} & e^{i k} \frac{1 - \delta}{2} \\
        e^{-i k} \frac{1 - \delta}{2} & e^{-i k} \frac{1 + \delta}{2}
    \end{pmatrix}
\end{align}
and has the characteristic polynomial
\begin{align*}
    \chi(\lambda) = \lambda^{2} - (e^{i k} + e^{-i k}) \frac{1 + \delta}{2} \lambda + \delta.
\end{align*}
Inserting its coefficients into \eqref{equation:FibonacciCoefficient} yields
\begin{align*}
    f_{n} = \sum_{h = 0}^{n} (-\delta)^{\frac{n - h}{2}} \frac{\left(\frac{n + h}{2}\right)!}{h! \left(\frac{n - h}{2}\right)!} \left(\frac{1 + \delta}{2}\right)^{h} \sum_{g = 0}^{h} \binom{h}{g} e^{i (2g - h) k}.
\end{align*}
The application of the Fibonacci-Horner-decomposed $M_{k}^{n}$ to $\tilde{r}_{0}(k) = (\alpha, \beta)^{\text{T}}$ yields the result
\begin{align*}
    \tilde{r}_{n}(k) = \begin{pmatrix}
        f_{n} - f_{n - 1} e^{-i k} \frac{1 + \delta}{2} & f_{n - 1} e^{i k} \frac{1 - \delta}{2} \\
        f_{n - 1} e^{-i k} \frac{1 - \delta}{2} & f_{n} - f_{n - 1} e^{i k} \frac{1 + \delta}{2}
    \end{pmatrix} \begin{pmatrix}
        \alpha_{\vphantom{n}} \\
        \beta_{\vphantom{n}}
    \end{pmatrix}.
\end{align*}
Applying the analogue of the inverse Fourier transform \eqref{equation:InverseFourierTransform} then proves \autoref{lemma:ClassicalWalkClosedForm}.

\section{\label{section:Variance}Variance}

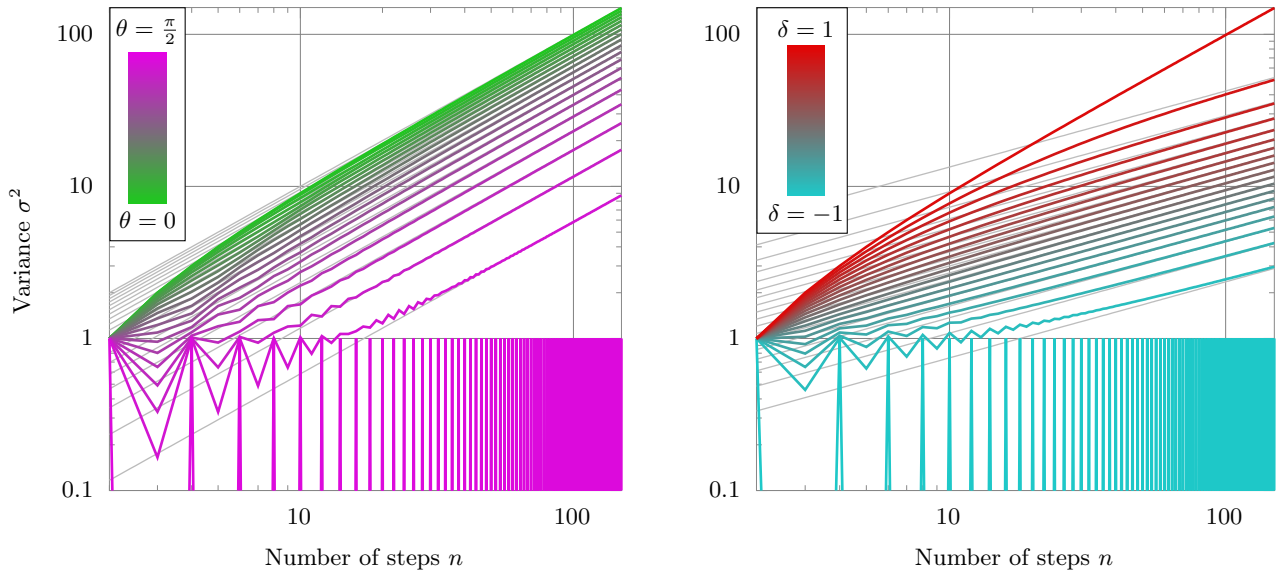
\begin{figure*}[!th]
    \begin{minipage}[t]{0.5\linewidth}
        \input{Quantum_Variance}
        \begin{tikzpicture}[overlay]
        \begin{scope}[xshift=-2.76cm,yshift=5.36cm]
        \draw[fill=white] (-0.25,-0.5) rectangle (0.75,2.6);
        \fill[shading=axis,left color = quantum_walk_pink,right color = quantum_walk_green,shading angle=0] (0,0) rectangle (0.5,2);
        \node[below] at (0.25,0) {$\theta = 0$};
        \node[above] at (0.25,2) {$\theta = \frac{\pi}{2}$};
        \end{scope}
    \end{tikzpicture}
    \end{minipage}\hfill
    \begin{minipage}[t]{0.5\linewidth}
        \input{Classical_Variance}
        \begin{tikzpicture}[overlay]
        \begin{scope}[xshift=-2.97cm,yshift=5.46cm]
        \draw[fill=white] (-0.4,-0.5) rectangle (0.8,2.5);
        \fill[shading=axis,left color = classical_walk_red,right color = classical_walk_blue,shading angle=0] (0,0) rectangle (0.5,2);
        \node[below] at (0.25,0) {$\delta = -1$};
        \node[above] at (0.25,2) {$\delta = 1$};
        \end{scope}
    \end{tikzpicture}
    \end{minipage}
    \caption{\label{figure:Variances}
        Variances of probability distributions after $n$ steps.
        The left plot shows the spatial variances of symmetric quantum walks with different values for the parameter $\theta \in [0, \pi / 2]$.
        The edge case $\theta = \pi / 2$ is the only case which does not produce asymptotically quadratically scaling variances.
        Its variances oscillate between $0$ (for even $n$) and $1$ (for odd $n$).
        For all other $\theta$-values, even for $\theta = 0$, the variances quickly assume their asymptotic quadratic scaling in $n$ (grey lines).
        In comparison, the right plot shows the spatial variances of symmetric correlated classical random walks with different values for the correlation coefficient $\delta \in [-1, 1]$.
        The edge case $\delta = -1$ produces the same probability distributions, hence the same variances, as the quantum walk for $\theta = \pi / 2$.
        The other edge case $\delta = 1$ yields the same probability distributions, hence the same variances, as the quantum walk for $\theta = 0$ and is the only case in which the variance increases quadratically with $n$.
        In all other cases, the variances quickly assume their asymptotic linear scaling in $n$ (grey lines).
    }
\end{figure*}

The arguably most striking characteristic of the Hadamard walk is its variance of order $\Omega(n^{2})$ \cite{Ambainis2001OneDimensionalQuantumWalks}.
In comparison, the (symmetric) uncorrelated classical random walk admits a scaling of $\Theta(n)$ in its variance.
We can also study the behaviour of the variance for finite $n$ thanks to the explicit formula for a non-trivial quantum walk's variance derived by Konno in \cite[Corollary 5]{Konno2005ANewTypeOfLimitTheoremsForTheOneDimensionalQuantumRandomWalk}:
\begin{align*}
\begin{split}
    \sigma^{2}\hspace*{-1pt} =\hspace*{-1pt} \abs{a}^{2 (n - 1)}\hspace*{-3pt} \left[n^{2}\hspace*{-2pt} +\hspace*{-4pt} \sum_{j\vphantom{k} = 1}^{\left\lceil\hspace*{-1pt}\frac{n - 1}{2}\hspace*{-1pt}\right\rceil}\hspace*{-5pt} \sum_{k, \ell = 1}^{j} \hspace*{-4pt} \left(\hspace*{-3pt}-\frac{\vert b\vert^{2}}{\abs{a}^{2}}\right)^{\hspace*{-3pt}k + \ell}\hspace*{-2pt} \frac{(n - 2 j)^{2} \kappa_{k, \ell, n, j}}{k \ell} \right] \\
    \text{with } \kappa_{k, \ell, n, j} \coloneqq \binom{k\hspace*{-1pt} -\hspace*{-1pt} 1}{\gamma\hspace*{-1pt} -\hspace*{-1pt} 1} \binom{k\hspace*{-1pt} -\hspace*{-1pt} 1}{\delta\hspace*{-1pt} -\hspace*{-1pt} 1} \binom{n\hspace*{-1pt} -\hspace*{-1pt} k\hspace*{-1pt} -\hspace*{-1pt} 1}{\gamma\hspace*{-1pt} -\hspace*{-1pt} 1} \binom{n\hspace*{-1pt} -\hspace*{-1pt} k\hspace*{-1pt} -\hspace*{-1pt} 1}{\delta\hspace*{-1pt} -\hspace*{-1pt} 1}.
\end{split}
\end{align*}
Interestingly, the variance (like all even moments) does not depend on the initial coin state's parameters $\alpha$ and $\beta$.
In fact, it only depends on $\abs{a}^{2}$ and $\lvert b\rvert^{2}$, hence only on the Hopf parameter $\theta$.
This means, for every coin setup $(C, \ket{\gamma})$ there exists another (usually non-equivalent) coin setup $(\tilde{C}, \ket{\tilde{\gamma}})$, necessarily with the same value for $\theta$, which induces a spatially symmetric quantum walk such that the quantum walk induced by $(C, \ket{\gamma})$ and the symmetric one induced by $(\tilde{C}, \ket{\tilde{\gamma}})$ give rise to probability distributions with the same variances after each step.

Most importantly, this formula shows that the quadratic asymptotic scaling of the variance with $n$ even holds true for non-trivial quantum walks with arbitrary coin setup.
The two trivial cases $\theta \in \{0, \pi / 2\}$ have to be analysed separately.

First, let $\theta = 0$.
By \autoref{theorem:ClassificationSymmetricWalks}, we may restrict our analysis to the coin $C = \upup + \downdown$ (i.e.\ the identity) and the coin state $\ket{\gamma} = (\up + i \down) / \sqrt{2}$.
The dynamics are, of course, trivial and we obtain
\begin{align*}
    \ket{\psi_{n}} = \frac{1}{\sqrt{2}} (i\ket{-n} \otimes \down + \ket{n} \otimes \up)
\end{align*}
with induced spatial probability distributions $p_{(C, \ket{\gamma})}(j, n) = (\delta_{j, -n} + \delta_{j, n}) / 2$.
These distributions clearly have a variance of $\sigma^{2} = n^{2}$, that is they also follow the asymptotic quadratic trend of the non-trivial quantum walks.

Second, let $\theta = \pi / 2$.
Again by \autoref{theorem:ClassificationSymmetricWalks}, we may restrict to the coin $C = \ket{\uparrow}\hspace*{-3pt}\bra{\downarrow} + \ket{\downarrow}\hspace*{-3pt}\bra{\uparrow}$ (i.e.\ the Pauli-X gate) and the coin state $\ket{\gamma} = (\up + i \down) / \sqrt{2}$.
Either with the aid of \autoref{lemma:QuantumWalkClosedForm} or by direct calculation, one verifies that
\begin{align*}
    \ket{\psi_{n}} = \frac{1}{\sqrt{2}} \begin{cases}
        \ket{0} \otimes (\up + i \down), & \text{if } n \text{ even}, \\
        (\ket{-1} \otimes \down + i \ket{1} \otimes \up), & \text{if } n \text{ odd},
    \end{cases}
\end{align*}
resulting in spatial probability distributions of the form
\begin{align*}
    p_{(C, \ket{\gamma})}(j, n) = \begin{cases}
        \delta_{j, 0}, & \text{if } n \text{ even}, \\
        \delta_{\abs{j}, 1} / 2, & \text{if } n \text{ odd}.
    \end{cases}
\end{align*}
The variances of these very simple probability distributions oscillate between $0$ (for even $n$) and $1$ (for odd $n$), i.e.\ they do not follow the asymptotic quadratic trend of the $\theta = 0$ case and the non-trivial quantum walks.

In comparison, Gillis derives the following expression for the variance of the correlated classical random walk with correlation coefficient $\delta \in [-1, 1)$ and initial parameters $\alpha = \beta = 1 / 2$ (compare \cite[3$\cdot$10]{Gillis1955CorrelatedRandomWalk}):
\begin{align*}
    \sigma^{2} = \frac{1 + \delta}{1 - \delta} n - \frac{2 \delta (1 - \delta^{n})}{(1 - \delta)^{2}}.
\end{align*}
Except for the completely (anti-)correlated case, the asymptotic scaling of the variance is linear in $n$.
For $\delta = \pm 1$, the resulting random walk readily produces the exact same probability distributions as the trivial quantum walks.
More specifically, $\delta = -1$ corresponds to the case where $\theta = \pi / 2$, and $\delta = 1$ to the case where $\theta = 0$.
Therefore, the completely correlated case of $\delta = 1$ is the only classical case in which $\sigma^{2}$ scales quadratically in $n$.
\autoref{figure:Variances} depicts the variance of the spatial probability distributions induced by symmetric quantum walks and correlated classical random walks for several $\theta$- and $\delta$-values, respectively.

\section{\label{section:LimitingDistributions}Limiting distributions}

\begin{figure*}[!th]
    \begin{minipage}[t]{0.5\linewidth}
        \input{Quantum_Distribution_Symmetric_Cut}
        \begin{tikzpicture}[overlay]
        \begin{scope}[xshift=-2.96cm,yshift=5.36cm]
        \draw[fill=white] (-0.25,-0.5) rectangle (0.75,2.6);
        \fill[shading=axis,left color = quantum_walk_pink,right color = quantum_walk_green,shading angle=0] (0,0) rectangle (0.5,2);
        \node[below] at (0.25,0) {$\theta = 0$};
        \node[above] at (0.25,2) {$\theta = \frac{\pi}{2}$};
        \end{scope}
    \end{tikzpicture}
    \end{minipage}\hfill
    \begin{minipage}[t]{0.5\linewidth}
        \input{Classical_Distribution}
        \begin{tikzpicture}[overlay]
        \begin{scope}[xshift=-2.97cm,yshift=5.46cm]
        \draw[fill=white] (-0.4,-0.5) rectangle (0.8,2.5);
        \fill[shading=axis,left color = classical_walk_red,right color = classical_walk_blue,shading angle=0] (0,0) rectangle (0.5,2);
        \node[below] at (0.25,0) {$\delta = -1$};
        \node[above] at (0.25,2) {$\delta = 1$};
        \end{scope}
    \end{tikzpicture} 
    \end{minipage} 
    \caption{\label{figure:SymmetricLimitingDistributions}
        Symmetric limiting distributions.
        The left plot shows the the limiting distribution densities $f_{\mathcal{C}}$ for coin setups inducing symmetric quantum walks with varying values for $\theta \in (0, \pi / 2)$.
        The distribution remains bimodal throughout the entire range of $\theta$-values with peaks at the boundary points $x = \pm \cos(\theta)$ after which it immediately drops to $0$.
        The distribution's spread decreases with increasing value of $\theta$.
        The right plot shows the $n^{-1 / 2}$-normalised limiting Gaussian densities for symmetric correlated classical random walks with varying values for $\delta \in (-1, 1)$.
        While the mean remains at $0$, the variance gradually increases with increasing value for $\delta$.
    }
\end{figure*}
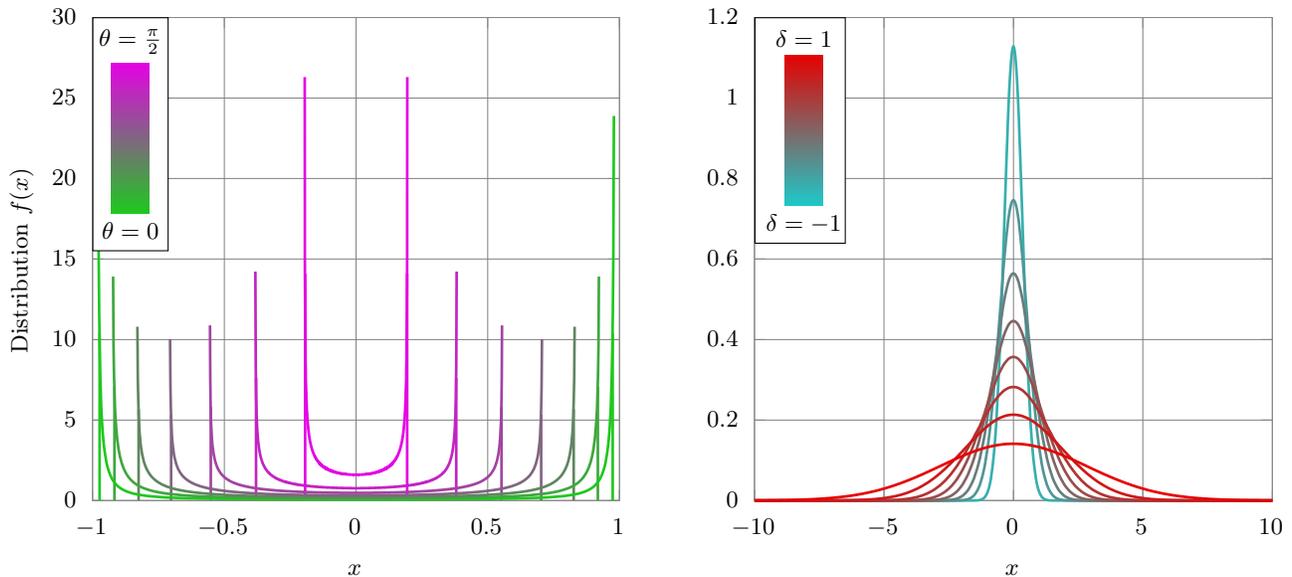

Recall from the proof of \autoref{theorem:ClassificationLimitingDistribution} the form of the density function for the limiting distributions of the sequence $p_{\mathcal{C}}(\defdot, n) / n$.
Further, following the proof of \autoref{theorem:ClassificationSymmetricWalks}, we already know that $\mathcal{C}$ induces a symmetric quantum walk if and only if $\abs{\alpha} = \vert\beta\vert^{2} = 1 / 2$ and $a \alpha \overline{b \beta} + \overline{a \alpha} b \beta = 0$.
That is, $\mathcal{C}$ inducing a symmetric quantum walk implies that $\lambda_{\mathcal{C}} = 0$ which, in turn, lifts all dependencies of $f_{\mathcal{C}}$ except the one on $\abs{a}^{2}$.
In this case, the limiting distribution therefore only depends on the \hyperref[equation:HopfCoordinates]{Hopf coordinate} $\theta$.
Furthermore, this dependency readily is injective which means the parametrisation \eqref{equation:SymmetricClassificationParametrisation} (excluding the boundary points $\theta = 0$ and $\theta = \pi / 2$) also captures all limiting distributions bijectively.

Now, one may be tempted to conjecture that bijectively covering all the limiting distributions for the general (not necessarily symmetric) quantum walk might also yield a bijective parametrisation of all the quantum walks modulo distributional equivalence for finite $n$.
While it still holds true that two distributionally equivalent coin setups also have the same limiting distribution, i.e., they are also asymptotically distributionally equivalent coin setups, 
the opposite does not hold as soon as we drop the symmetry requirement.

As a concrete example, take two coin setups $\mathcal{C}_{1}$ and $\mathcal{C}_{2}$ whose coins have the same non-zero value for $\theta$ and whose coin states both have $\xi = 0$.
For $\mathcal{C} \in \{\mathcal{C}_{1}, \mathcal{C}_{2}\}$, $\lambda_{\mathcal{C}}$ simplifies to
\begin{align*}
    \lambda_{\mathcal{C}} = \sqrt{1 + \tan^{2}\hspace*{-1pt}(\theta)} \cos(2 \varphi - \theta).
\end{align*}
Choosing the $\varphi$ parameters for the two coin states such that $\varphi_{1} + \varphi_{2} = \theta$ then implies
\begin{align*}
    \cos(2 \varphi - \theta) = \cos(2 (\theta - \varphi_{2}) - \theta) = \cos(\theta - 2 \varphi_{2}),
\end{align*}
which, since $\cos$ is an even function yields that $\lambda_{\mathcal{C}_{1}} = \lambda_{\mathcal{C}_{2}}$.
That is, any such pair of coin setup has the same values for $\theta$ and $\lambda_{\mathcal{C}}$ and hence induces the same probability distribution.
However, a concrete calculation for, e.g., $\theta = 1.2$, $\varphi_{1} = 0.2$, and $\varphi_{2} = 1.0$ shows that the probability distributions for finite $n$ (e.g., $n = 2$) do not coincide.

In contrast to the rather complicated form of $f_{\mathcal{C}}$, even for symmetric setups, the correlated classical case has considerably simpler limiting distribution.
As shown in \cite[645]{Gillis1955CorrelatedRandomWalk}, random walks with correlation coefficient $\delta \in (-1, 1)$ induce probability distributions whose limiting distribution is Gaussian with zero mean (i.e., spatially symmetric) and a variance of
\begin{align*}
    \sigma^{2} = \frac{n (1 + \delta)}{1 - \delta}.
\end{align*}

\autoref{figure:SymmetricLimitingDistributions} depicts the limiting distributions for the symmetric quantum and classical case for several $\theta$- and $\delta$-values, respectively.
The quantum case produces bimodal limiting distributions whose densities are minimal in the centre $x = 0$ and monotonically increasing toward $x = \pm \cos(\theta)$, after which they drop to $0$.
In contrast, the classical case's Gaussian limiting distributions maintain their maximum at the centre $x = 0$ and quickly decay on both sides.

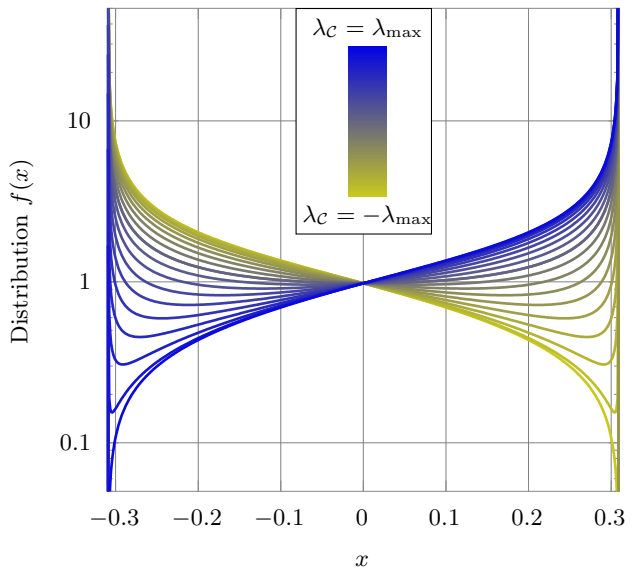
\begin{figure}
    \input{Quantum_Distribution}
    \begin{tikzpicture}[overlay]
    \begin{scope}[xshift=0.35cm,yshift=5.46cm]
        \draw[fill=white] (-0.7,-0.5) rectangle (1.1,2.5);
        \fill[shading=axis,left color = symmetry_walk_yellow,right color = symmetry_walk_blue,shading angle=0] (0,0) rectangle (0.5,2);
        \node[below] at (0.25,0) {$\lambda_{\mathcal{C}} = -\lambda_{\max}$};
        \node[above] at (0.25,2) {$\lambda_{\mathcal{C}} = \lambda_{\max}$};
        \end{scope}
    \end{tikzpicture}
    \caption{\label{figure:AsymmetricLimitingDistributions}
        Limiting distribution densities $f_{\mathcal{C}}$ for coin setups with $\theta = 0.4 \pi$ and various combinations of $\xi$- and $\varphi$-values to produce $\lambda_{\mathcal{C}}$-values between $\pm \lambda_{\max}$, where $\lambda_{\max} = \sqrt{1 + \tan^{2}(\theta)}$.
        This semilog plot shows how larger absolute values for $\lambda_{\mathcal{C}}$ lead to more asymmetric distributions, while $\lambda_{\mathcal{C}} = 0$ yields a symmetric distribution.
    }
\end{figure}

\autoref{figure:AsymmetricLimitingDistributions} displays asymmetric limiting distributions for quantum walks for a fixed value of $\theta$, but with varying $\lambda_{\mathcal{C}} \in [-\sqrt{1 + \tan^{2}(\theta)}, \sqrt{1 + \tan^{2}(\theta)}]$.
One can directly observe how the $\lambda_{\mathcal{C}}$-parameter controls the asymmetry of the distribution.
For $\lambda_{\mathcal{C}} = 0$ we recover the symmetric case.

\section{\label{section:Conclusion}Conclusion}

In this work, we provided a comprehensive classification of one-dimensional coined quantum walks in terms of the probability distributions they induce.
Focusing on the spatial symmetry of these distributions, we identified all initial coin states that yield symmetric walks for arbitrary quantum coins and constructed a bijective parametrisation of the equivalence classes of such walks.
Extending beyond the symmetric case, we also derived a surjective parametrisation of all coined quantum walks modulo distributional equivalence and a bijective parametrisation of all possible limiting distributions.

We further corrected and refined recent results on closed-form expressions for the walker's amplitudes, resolving a discrepancy in earlier calculations, and extended the methodology to the classically correlated random walk.
This enabled a unified treatment of both quantum and classical dynamics within the same analytical framework.

The variance analysis and plots confirmed that quadratic spreading is a robust feature of non-trivial quantum walks, independent of symmetry or initial coin state, with only one trivial case deviating from this behaviour.
In contrast, correlated classical walks exhibit linear spreading except at the extremal points of full correlation, thereby highlighting a fundamental distinction in their dynamical regimes.

Furthermore, we constructed a class of counter examples showcasing that two quantum walks can have the same limiting distribution even if their induced distributions after finitely many steps differ.
We were able to verify the validity of these counter examples with the corrected closed-form expression for the quantum walker's amplitudes.
Via additional graphical representations of symmetric limiting distributions for quantum and correlated classical random walks, we drew further comparisons between both regimes, highlighting bimodality and minimal concentration at the centre as distinct quantum features.
In an additional plot for asymmetric limiting distributions of quantum walks, we qualitatively studied attainable shapes of quantum limiting distributions.

Altogether, our results deepen the understanding of structural features in coined quantum walks and establish a foundation for further analytical comparisons with classical stochastic processes, particularly in contexts where symmetry, coherence, or controllable correlations play a pivotal role.
For example, our method lays the foundation for comparing quantum walks to more general Markov chains than those given by correlated random walks.
This extended classical setting would also allow for non-symmetric limiting distributions, something that is unattainable with correlated random walks.

The most interesting task which still remains open is to give a bijective parametrisation of all coined quantum walks modulo distributional equivalence.
Here, our surjective parametrisation might yield ``outer bounds'' on the necessary parameter domains while our bijective parametrisation of all limiting distributions might constitute some sort of ``inner bounds'' on the domains.

\vspace*{-0.5cm}

\begin{acknowledgments}

\vspace*{-0.3cm}

LB acknowledges financial support by the Quantum Valley Lower Saxony. LH acknowledges financial support by the Alexander von Humboldt Foundation.

\noindent\textbf{Data and code availability statement.}
The depicted data as well as the source code are publicly available on GitHub: \url{https://github.com/HANTLUK/QuantumRandomWalks}.

\end{acknowledgments}

\bibliographystyle{apsrev4-2}
\bibliography{main.bib}

\twocolumngrid

\onecolumngrid

\end{document}

%% file: Quantum_Variance.tex
% GNUPLOT: LaTeX picture with Postscript
\begingroup
  \makeatletter
  \providecommand\color[2][]{%
    \GenericError{(gnuplot) \space\space\space\@spaces}{%
      Package color not loaded in conjunction with
      terminal option `colourtext'%
    }{See the gnuplot documentation for explanation.%
    }{Either use 'blacktext' in gnuplot or load the package
      color.sty in LaTeX.}%
    \renewcommand\color[2][]{}%
  }%
  \providecommand\includegraphics[2][]{%
    \GenericError{(gnuplot) \space\space\space\@spaces}{%
      Package graphicx or graphics not loaded%
    }{See the gnuplot documentation for explanation.%
    }{The gnuplot epslatex terminal needs graphicx.sty or graphics.sty.}%
    \renewcommand\includegraphics[2][]{}%
  }%
  \providecommand\rotatebox[2]{#2}%
  \@ifundefined{ifGPcolor}{%
    \newif\ifGPcolor
    \GPcolortrue
  }{}%
  \@ifundefined{ifGPblacktext}{%
    \newif\ifGPblacktext
    \GPblacktextfalse
  }{}%
  % define a \g@addto@macro without @ in the name:
  \let\gplgaddtomacro\g@addto@macro
  % define empty templates for all commands taking text:
  \gdef\gplbacktext{}%
  \gdef\gplfronttext{}%
  \makeatother
  \ifGPblacktext
    % no textcolor at all
    \def\colorrgb#1{}%
    \def\colorgray#1{}%
  \else
    % gray or color?
    \ifGPcolor
      \def\colorrgb#1{\color[rgb]{#1}}%
      \def\colorgray#1{\color[gray]{#1}}%
      \expandafter\def\csname LTw\endcsname{\color{white}}%
      \expandafter\def\csname LTb\endcsname{\color{black}}%
      \expandafter\def\csname LTa\endcsname{\color{black}}%
      \expandafter\def\csname LT0\endcsname{\color[rgb]{1,0,0}}%
      \expandafter\def\csname LT1\endcsname{\color[rgb]{0,1,0}}%
      \expandafter\def\csname LT2\endcsname{\color[rgb]{0,0,1}}%
      \expandafter\def\csname LT3\endcsname{\color[rgb]{1,0,1}}%
      \expandafter\def\csname LT4\endcsname{\color[rgb]{0,1,1}}%
      \expandafter\def\csname LT5\endcsname{\color[rgb]{1,1,0}}%
      \expandafter\def\csname LT6\endcsname{\color[rgb]{0,0,0}}%
      \expandafter\def\csname LT7\endcsname{\color[rgb]{1,0.3,0}}%
      \expandafter\def\csname LT8\endcsname{\color[rgb]{0.5,0.5,0.5}}%
    \else
      % gray
      \def\colorrgb#1{\color{black}}%
      \def\colorgray#1{\color[gray]{#1}}%
      \expandafter\def\csname LTw\endcsname{\color{white}}%
      \expandafter\def\csname LTb\endcsname{\color{black}}%
      \expandafter\def\csname LTa\endcsname{\color{black}}%
      \expandafter\def\csname LT0\endcsname{\color{black}}%
      \expandafter\def\csname LT1\endcsname{\color{black}}%
      \expandafter\def\csname LT2\endcsname{\color{black}}%
      \expandafter\def\csname LT3\endcsname{\color{black}}%
      \expandafter\def\csname LT4\endcsname{\color{black}}%
      \expandafter\def\csname LT5\endcsname{\color{black}}%
      \expandafter\def\csname LT6\endcsname{\color{black}}%
      \expandafter\def\csname LT7\endcsname{\color{black}}%
      \expandafter\def\csname LT8\endcsname{\color{black}}%
    \fi
  \fi
    \setlength{\unitlength}{0.0500bp}%
    \ifx\gptboxheight\undefined%
      \newlength{\gptboxheight}%
      \newlength{\gptboxwidth}%
      \newsavebox{\gptboxtext}%
    \fi%
    \setlength{\fboxrule}{0.5pt}%
    \setlength{\fboxsep}{1pt}%
    \definecolor{tbcol}{rgb}{1,1,1}%
\begin{picture}(5260.00,4520.00)%
    \gplgaddtomacro\gplbacktext{%
      \colorrgb{0.00,0.00,0.00}%%
      \put(708,652){\makebox(0,0)[r]{\strut{}$0.1$}}%
      \colorrgb{0.00,0.00,0.00}%%
      \put(708,1799){\makebox(0,0)[r]{\strut{}$1$}}%
      \colorrgb{0.00,0.00,0.00}%%
      \put(708,2946){\makebox(0,0)[r]{\strut{}$10$}}%
      \colorrgb{0.00,0.00,0.00}%%
      \put(708,4093){\makebox(0,0)[r]{\strut{}$100$}}%
      \colorrgb{0.00,0.00,0.00}%%
      \put(2259,448){\makebox(0,0){\strut{}$10$}}%
      \colorrgb{0.00,0.00,0.00}%%
      \put(4317,448){\makebox(0,0){\strut{}$100$}}%
    }%
    \gplgaddtomacro\gplfronttext{%
      \csname LTb\endcsname%%
      \put(186,2473){\rotatebox{-270}{\makebox(0,0){\strut{}Variance $\sigma^2$}}}%
      \csname LTb\endcsname%%
      \put(2749,142){\makebox(0,0){\strut{}Number of steps $n$}}%
    }%
    \gplbacktext
    \put(0,0){\includegraphics[width={263.00bp},height={226.00bp}]{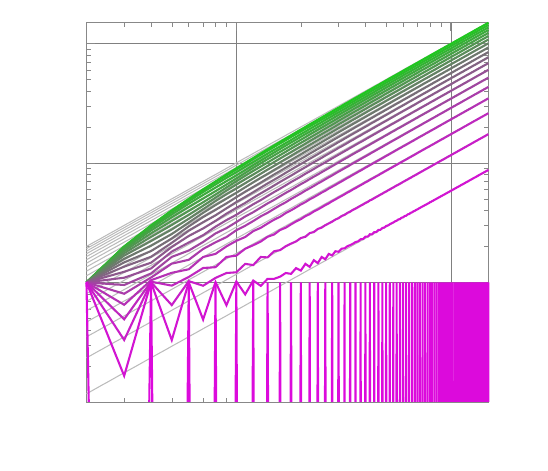}}%
    \gplfronttext
  \end{picture}%
\endgroup

%% file: Classical_Variance.tex
% GNUPLOT: LaTeX picture with Postscript
\begingroup
  \makeatletter
  \providecommand\color[2][]{%
    \GenericError{(gnuplot) \space\space\space\@spaces}{%
      Package color not loaded in conjunction with
      terminal option `colourtext'%
    }{See the gnuplot documentation for explanation.%
    }{Either use 'blacktext' in gnuplot or load the package
      color.sty in LaTeX.}%
    \renewcommand\color[2][]{}%
  }%
  \providecommand\includegraphics[2][]{%
    \GenericError{(gnuplot) \space\space\space\@spaces}{%
      Package graphicx or graphics not loaded%
    }{See the gnuplot documentation for explanation.%
    }{The gnuplot epslatex terminal needs graphicx.sty or graphics.sty.}%
    \renewcommand\includegraphics[2][]{}%
  }%
  \providecommand\rotatebox[2]{#2}%
  \@ifundefined{ifGPcolor}{%
    \newif\ifGPcolor
    \GPcolortrue
  }{}%
  \@ifundefined{ifGPblacktext}{%
    \newif\ifGPblacktext
    \GPblacktextfalse
  }{}%
  % define a \g@addto@macro without @ in the name:
  \let\gplgaddtomacro\g@addto@macro
  % define empty templates for all commands taking text:
  \gdef\gplbacktext{}%
  \gdef\gplfronttext{}%
  \makeatother
  \ifGPblacktext
    % no textcolor at all
    \def\colorrgb#1{}%
    \def\colorgray#1{}%
  \else
    % gray or color?
    \ifGPcolor
      \def\colorrgb#1{\color[rgb]{#1}}%
      \def\colorgray#1{\color[gray]{#1}}%
      \expandafter\def\csname LTw\endcsname{\color{white}}%
      \expandafter\def\csname LTb\endcsname{\color{black}}%
      \expandafter\def\csname LTa\endcsname{\color{black}}%
      \expandafter\def\csname LT0\endcsname{\color[rgb]{1,0,0}}%
      \expandafter\def\csname LT1\endcsname{\color[rgb]{0,1,0}}%
      \expandafter\def\csname LT2\endcsname{\color[rgb]{0,0,1}}%
      \expandafter\def\csname LT3\endcsname{\color[rgb]{1,0,1}}%
      \expandafter\def\csname LT4\endcsname{\color[rgb]{0,1,1}}%
      \expandafter\def\csname LT5\endcsname{\color[rgb]{1,1,0}}%
      \expandafter\def\csname LT6\endcsname{\color[rgb]{0,0,0}}%
      \expandafter\def\csname LT7\endcsname{\color[rgb]{1,0.3,0}}%
      \expandafter\def\csname LT8\endcsname{\color[rgb]{0.5,0.5,0.5}}%
    \else
      % gray
      \def\colorrgb#1{\color{black}}%
      \def\colorgray#1{\color[gray]{#1}}%
      \expandafter\def\csname LTw\endcsname{\color{white}}%
      \expandafter\def\csname LTb\endcsname{\color{black}}%
      \expandafter\def\csname LTa\endcsname{\color{black}}%
      \expandafter\def\csname LT0\endcsname{\color{black}}%
      \expandafter\def\csname LT1\endcsname{\color{black}}%
      \expandafter\def\csname LT2\endcsname{\color{black}}%
      \expandafter\def\csname LT3\endcsname{\color{black}}%
      \expandafter\def\csname LT4\endcsname{\color{black}}%
      \expandafter\def\csname LT5\endcsname{\color{black}}%
      \expandafter\def\csname LT6\endcsname{\color{black}}%
      \expandafter\def\csname LT7\endcsname{\color{black}}%
      \expandafter\def\csname LT8\endcsname{\color{black}}%
    \fi
  \fi
    \setlength{\unitlength}{0.0500bp}%
    \ifx\gptboxheight\undefined%
      \newlength{\gptboxheight}%
      \newlength{\gptboxwidth}%
      \newsavebox{\gptboxtext}%
    \fi%
    \setlength{\fboxrule}{0.5pt}%
    \setlength{\fboxsep}{1pt}%
    \definecolor{tbcol}{rgb}{1,1,1}%
\begin{picture}(5100.00,4520.00)%
    \gplgaddtomacro\gplbacktext{%
      \colorrgb{0.00,0.00,0.00}%%
      \put(504,652){\makebox(0,0)[r]{\strut{}$0.1$}}%
      \colorrgb{0.00,0.00,0.00}%%
      \put(504,1799){\makebox(0,0)[r]{\strut{}$1$}}%
      \colorrgb{0.00,0.00,0.00}%%
      \put(504,2946){\makebox(0,0)[r]{\strut{}$10$}}%
      \colorrgb{0.00,0.00,0.00}%%
      \put(504,4093){\makebox(0,0)[r]{\strut{}$100$}}%
      \colorrgb{0.00,0.00,0.00}%%
      \put(2071,448){\makebox(0,0){\strut{}$10$}}%
      \colorrgb{0.00,0.00,0.00}%%
      \put(4152,448){\makebox(0,0){\strut{}$100$}}%
    }%
    \gplgaddtomacro\gplfronttext{%
      \csname LTb\endcsname%%
      \put(84,2473){\rotatebox{-270}{\makebox(0,0){\strut{}}}}%
      \csname LTb\endcsname%%
      \put(2567,142){\makebox(0,0){\strut{}Number of steps $n$}}%
      \csname LTb\endcsname%%
      \put(2567,4295){\makebox(0,0){\strut{}}}%
    }%
    \gplbacktext
    \put(0,0){\includegraphics[width={255.00bp},height={226.00bp}]{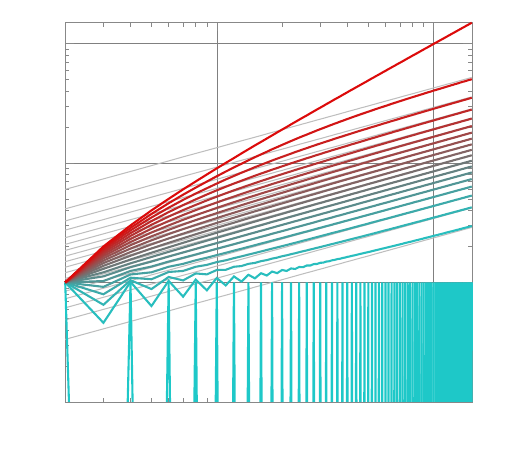}}%
    \gplfronttext
  \end{picture}%
\endgroup

%% file: Quantum_Distribution_Symmetric_Cut.tex
% GNUPLOT: LaTeX picture with Postscript
\begingroup
  \makeatletter
  \providecommand\color[2][]{%
    \GenericError{(gnuplot) \space\space\space\@spaces}{%
      Package color not loaded in conjunction with
      terminal option `colourtext'%
    }{See the gnuplot documentation for explanation.%
    }{Either use 'blacktext' in gnuplot or load the package
      color.sty in LaTeX.}%
    \renewcommand\color[2][]{}%
  }%
  \providecommand\includegraphics[2][]{%
    \GenericError{(gnuplot) \space\space\space\@spaces}{%
      Package graphicx or graphics not loaded%
    }{See the gnuplot documentation for explanation.%
    }{The gnuplot epslatex terminal needs graphicx.sty or graphics.sty.}%
    \renewcommand\includegraphics[2][]{}%
  }%
  \providecommand\rotatebox[2]{#2}%
  \@ifundefined{ifGPcolor}{%
    \newif\ifGPcolor
    \GPcolortrue
  }{}%
  \@ifundefined{ifGPblacktext}{%
    \newif\ifGPblacktext
    \GPblacktextfalse
  }{}%
  % define a \g@addto@macro without @ in the name:
  \let\gplgaddtomacro\g@addto@macro
  % define empty templates for all commands taking text:
  \gdef\gplbacktext{}%
  \gdef\gplfronttext{}%
  \makeatother
  \ifGPblacktext
    % no textcolor at all
    \def\colorrgb#1{}%
    \def\colorgray#1{}%
  \else
    % gray or color?
    \ifGPcolor
      \def\colorrgb#1{\color[rgb]{#1}}%
      \def\colorgray#1{\color[gray]{#1}}%
      \expandafter\def\csname LTw\endcsname{\color{white}}%
      \expandafter\def\csname LTb\endcsname{\color{black}}%
      \expandafter\def\csname LTa\endcsname{\color{black}}%
      \expandafter\def\csname LT0\endcsname{\color[rgb]{1,0,0}}%
      \expandafter\def\csname LT1\endcsname{\color[rgb]{0,1,0}}%
      \expandafter\def\csname LT2\endcsname{\color[rgb]{0,0,1}}%
      \expandafter\def\csname LT3\endcsname{\color[rgb]{1,0,1}}%
      \expandafter\def\csname LT4\endcsname{\color[rgb]{0,1,1}}%
      \expandafter\def\csname LT5\endcsname{\color[rgb]{1,1,0}}%
      \expandafter\def\csname LT6\endcsname{\color[rgb]{0,0,0}}%
      \expandafter\def\csname LT7\endcsname{\color[rgb]{1,0.3,0}}%
      \expandafter\def\csname LT8\endcsname{\color[rgb]{0.5,0.5,0.5}}%
    \else
      % gray
      \def\colorrgb#1{\color{black}}%
      \def\colorgray#1{\color[gray]{#1}}%
      \expandafter\def\csname LTw\endcsname{\color{white}}%
      \expandafter\def\csname LTb\endcsname{\color{black}}%
      \expandafter\def\csname LTa\endcsname{\color{black}}%
      \expandafter\def\csname LT0\endcsname{\color{black}}%
      \expandafter\def\csname LT1\endcsname{\color{black}}%
      \expandafter\def\csname LT2\endcsname{\color{black}}%
      \expandafter\def\csname LT3\endcsname{\color{black}}%
      \expandafter\def\csname LT4\endcsname{\color{black}}%
      \expandafter\def\csname LT5\endcsname{\color{black}}%
      \expandafter\def\csname LT6\endcsname{\color{black}}%
      \expandafter\def\csname LT7\endcsname{\color{black}}%
      \expandafter\def\csname LT8\endcsname{\color{black}}%
    \fi
  \fi
    \setlength{\unitlength}{0.0500bp}%
    \ifx\gptboxheight\undefined%
      \newlength{\gptboxheight}%
      \newlength{\gptboxwidth}%
      \newsavebox{\gptboxtext}%
    \fi%
    \setlength{\fboxrule}{0.5pt}%
    \setlength{\fboxsep}{1pt}%
    \definecolor{tbcol}{rgb}{1,1,1}%
\begin{picture}(5260.00,4520.00)%
    \gplgaddtomacro\gplbacktext{%
      \colorrgb{0.00,0.00,0.00}%%
      \put(596,652){\makebox(0,0)[r]{\strut{}$0$}}%
      \colorrgb{0.00,0.00,0.00}%%
      \put(596,1259){\makebox(0,0)[r]{\strut{}$5$}}%
      \colorrgb{0.00,0.00,0.00}%%
      \put(596,1866){\makebox(0,0)[r]{\strut{}$10$}}%
      \colorrgb{0.00,0.00,0.00}%%
      \put(596,2474){\makebox(0,0)[r]{\strut{}$15$}}%
      \colorrgb{0.00,0.00,0.00}%%
      \put(596,3081){\makebox(0,0)[r]{\strut{}$20$}}%
      \colorrgb{0.00,0.00,0.00}%%
      \put(596,3688){\makebox(0,0)[r]{\strut{}$25$}}%
      \colorrgb{0.00,0.00,0.00}%%
      \put(596,4295){\makebox(0,0)[r]{\strut{}$30$}}%
      \colorrgb{0.00,0.00,0.00}%%
      \put(708,448){\makebox(0,0){\strut{}$-1$}}%
      \colorrgb{0.00,0.00,0.00}%%
      \put(1701,448){\makebox(0,0){\strut{}$-0.5$}}%
      \colorrgb{0.00,0.00,0.00}%%
      \put(2694,448){\makebox(0,0){\strut{}$0$}}%
      \colorrgb{0.00,0.00,0.00}%%
      \put(3686,448){\makebox(0,0){\strut{}$0.5$}}%
      \colorrgb{0.00,0.00,0.00}%%
      \put(4679,448){\makebox(0,0){\strut{}$1$}}%
    }%
    \gplgaddtomacro\gplfronttext{%
      \csname LTb\endcsname%%
      \put(186,2473){\rotatebox{-270}{\makebox(0,0){\strut{}Distribution $f(x)$}}}%
      \csname LTb\endcsname%%
      \put(2693,142){\makebox(0,0){\strut{}$x$}}%
    }%
    \gplbacktext
    \put(0,0){\includegraphics[width={263.00bp},height={226.00bp}]{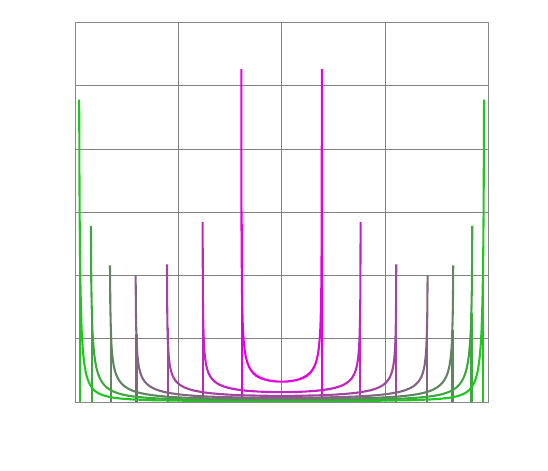}}%
    \gplfronttext
  \end{picture}%
\endgroup

%% file: Classical_Distribution.tex
% GNUPLOT: LaTeX picture with Postscript
\begingroup
  \makeatletter
  \providecommand\color[2][]{%
    \GenericError{(gnuplot) \space\space\space\@spaces}{%
      Package color not loaded in conjunction with
      terminal option `colourtext'%
    }{See the gnuplot documentation for explanation.%
    }{Either use 'blacktext' in gnuplot or load the package
      color.sty in LaTeX.}%
    \renewcommand\color[2][]{}%
  }%
  \providecommand\includegraphics[2][]{%
    \GenericError{(gnuplot) \space\space\space\@spaces}{%
      Package graphicx or graphics not loaded%
    }{See the gnuplot documentation for explanation.%
    }{The gnuplot epslatex terminal needs graphicx.sty or graphics.sty.}%
    \renewcommand\includegraphics[2][]{}%
  }%
  \providecommand\rotatebox[2]{#2}%
  \@ifundefined{ifGPcolor}{%
    \newif\ifGPcolor
    \GPcolortrue
  }{}%
  \@ifundefined{ifGPblacktext}{%
    \newif\ifGPblacktext
    \GPblacktextfalse
  }{}%
  % define a \g@addto@macro without @ in the name:
  \let\gplgaddtomacro\g@addto@macro
  % define empty templates for all commands taking text:
  \gdef\gplbacktext{}%
  \gdef\gplfronttext{}%
  \makeatother
  \ifGPblacktext
    % no textcolor at all
    \def\colorrgb#1{}%
    \def\colorgray#1{}%
  \else
    % gray or color?
    \ifGPcolor
      \def\colorrgb#1{\color[rgb]{#1}}%
      \def\colorgray#1{\color[gray]{#1}}%
      \expandafter\def\csname LTw\endcsname{\color{white}}%
      \expandafter\def\csname LTb\endcsname{\color{black}}%
      \expandafter\def\csname LTa\endcsname{\color{black}}%
      \expandafter\def\csname LT0\endcsname{\color[rgb]{1,0,0}}%
      \expandafter\def\csname LT1\endcsname{\color[rgb]{0,1,0}}%
      \expandafter\def\csname LT2\endcsname{\color[rgb]{0,0,1}}%
      \expandafter\def\csname LT3\endcsname{\color[rgb]{1,0,1}}%
      \expandafter\def\csname LT4\endcsname{\color[rgb]{0,1,1}}%
      \expandafter\def\csname LT5\endcsname{\color[rgb]{1,1,0}}%
      \expandafter\def\csname LT6\endcsname{\color[rgb]{0,0,0}}%
      \expandafter\def\csname LT7\endcsname{\color[rgb]{1,0.3,0}}%
      \expandafter\def\csname LT8\endcsname{\color[rgb]{0.5,0.5,0.5}}%
    \else
      % gray
      \def\colorrgb#1{\color{black}}%
      \def\colorgray#1{\color[gray]{#1}}%
      \expandafter\def\csname LTw\endcsname{\color{white}}%
      \expandafter\def\csname LTb\endcsname{\color{black}}%
      \expandafter\def\csname LTa\endcsname{\color{black}}%
      \expandafter\def\csname LT0\endcsname{\color{black}}%
      \expandafter\def\csname LT1\endcsname{\color{black}}%
      \expandafter\def\csname LT2\endcsname{\color{black}}%
      \expandafter\def\csname LT3\endcsname{\color{black}}%
      \expandafter\def\csname LT4\endcsname{\color{black}}%
      \expandafter\def\csname LT5\endcsname{\color{black}}%
      \expandafter\def\csname LT6\endcsname{\color{black}}%
      \expandafter\def\csname LT7\endcsname{\color{black}}%
      \expandafter\def\csname LT8\endcsname{\color{black}}%
    \fi
  \fi
    \setlength{\unitlength}{0.0500bp}%
    \ifx\gptboxheight\undefined%
      \newlength{\gptboxheight}%
      \newlength{\gptboxwidth}%
      \newsavebox{\gptboxtext}%
    \fi%
    \setlength{\fboxrule}{0.5pt}%
    \setlength{\fboxsep}{1pt}%
    \definecolor{tbcol}{rgb}{1,1,1}%
\begin{picture}(5100.00,4520.00)%
    \gplgaddtomacro\gplbacktext{%
      \colorrgb{0.00,0.00,0.00}%%
      \put(504,652){\makebox(0,0)[r]{\strut{}$0$}}%
      \colorrgb{0.00,0.00,0.00}%%
      \put(504,1259){\makebox(0,0)[r]{\strut{}$0.2$}}%
      \colorrgb{0.00,0.00,0.00}%%
      \put(504,1866){\makebox(0,0)[r]{\strut{}$0.4$}}%
      \colorrgb{0.00,0.00,0.00}%%
      \put(504,2474){\makebox(0,0)[r]{\strut{}$0.6$}}%
      \colorrgb{0.00,0.00,0.00}%%
      \put(504,3081){\makebox(0,0)[r]{\strut{}$0.8$}}%
      \colorrgb{0.00,0.00,0.00}%%
      \put(504,3688){\makebox(0,0)[r]{\strut{}$1$}}%
      \colorrgb{0.00,0.00,0.00}%%
      \put(504,4295){\makebox(0,0)[r]{\strut{}$1.2$}}%
      \colorrgb{0.00,0.00,0.00}%%
      \put(616,448){\makebox(0,0){\strut{}$-10$}}%
      \colorrgb{0.00,0.00,0.00}%%
      \put(1592,448){\makebox(0,0){\strut{}$-5$}}%
      \colorrgb{0.00,0.00,0.00}%%
      \put(2568,448){\makebox(0,0){\strut{}$0$}}%
      \colorrgb{0.00,0.00,0.00}%%
      \put(3543,448){\makebox(0,0){\strut{}$5$}}%
      \colorrgb{0.00,0.00,0.00}%%
      \put(4519,448){\makebox(0,0){\strut{}$10$}}%
    }%
    \gplgaddtomacro\gplfronttext{%
      \csname LTb\endcsname%%
      \put(84,2473){\rotatebox{-270}{\makebox(0,0){\strut{}}}}%
      \csname LTb\endcsname%%
      \put(2567,142){\makebox(0,0){\strut{}$x$}}%
      \csname LTb\endcsname%%
      \put(2567,4295){\makebox(0,0){\strut{}}}%
    }%
    \gplbacktext
    \put(0,0){\includegraphics[width={255.00bp},height={226.00bp}]{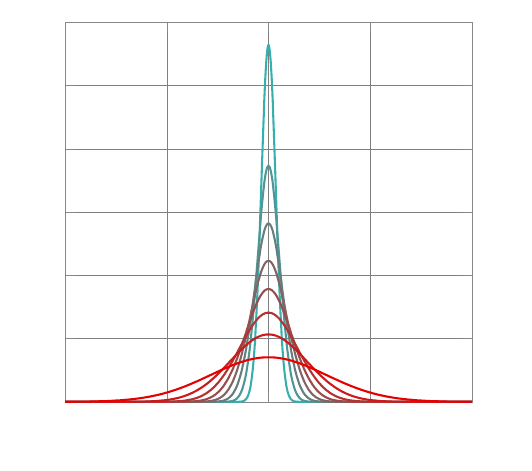}}%
    \gplfronttext
  \end{picture}%
\endgroup

%% file: Quantum_Distribution.tex
% GNUPLOT: LaTeX picture with Postscript
\begingroup
  \makeatletter
  \providecommand\color[2][]{%
    \GenericError{(gnuplot) \space\space\space\@spaces}{%
      Package color not loaded in conjunction with
      terminal option `colourtext'%
    }{See the gnuplot documentation for explanation.%
    }{Either use 'blacktext' in gnuplot or load the package
      color.sty in LaTeX.}%
    \renewcommand\color[2][]{}%
  }%
  \providecommand\includegraphics[2][]{%
    \GenericError{(gnuplot) \space\space\space\@spaces}{%
      Package graphicx or graphics not loaded%
    }{See the gnuplot documentation for explanation.%
    }{The gnuplot epslatex terminal needs graphicx.sty or graphics.sty.}%
    \renewcommand\includegraphics[2][]{}%
  }%
  \providecommand\rotatebox[2]{#2}%
  \@ifundefined{ifGPcolor}{%
    \newif\ifGPcolor
    \GPcolortrue
  }{}%
  \@ifundefined{ifGPblacktext}{%
    \newif\ifGPblacktext
    \GPblacktextfalse
  }{}%
  % define a \g@addto@macro without @ in the name:
  \let\gplgaddtomacro\g@addto@macro
  % define empty templates for all commands taking text:
  \gdef\gplbacktext{}%
  \gdef\gplfronttext{}%
  \makeatother
  \ifGPblacktext
    % no textcolor at all
    \def\colorrgb#1{}%
    \def\colorgray#1{}%
  \else
    % gray or color?
    \ifGPcolor
      \def\colorrgb#1{\color[rgb]{#1}}%
      \def\colorgray#1{\color[gray]{#1}}%
      \expandafter\def\csname LTw\endcsname{\color{white}}%
      \expandafter\def\csname LTb\endcsname{\color{black}}%
      \expandafter\def\csname LTa\endcsname{\color{black}}%
      \expandafter\def\csname LT0\endcsname{\color[rgb]{1,0,0}}%
      \expandafter\def\csname LT1\endcsname{\color[rgb]{0,1,0}}%
      \expandafter\def\csname LT2\endcsname{\color[rgb]{0,0,1}}%
      \expandafter\def\csname LT3\endcsname{\color[rgb]{1,0,1}}%
      \expandafter\def\csname LT4\endcsname{\color[rgb]{0,1,1}}%
      \expandafter\def\csname LT5\endcsname{\color[rgb]{1,1,0}}%
      \expandafter\def\csname LT6\endcsname{\color[rgb]{0,0,0}}%
      \expandafter\def\csname LT7\endcsname{\color[rgb]{1,0.3,0}}%
      \expandafter\def\csname LT8\endcsname{\color[rgb]{0.5,0.5,0.5}}%
    \else
      % gray
      \def\colorrgb#1{\color{black}}%
      \def\colorgray#1{\color[gray]{#1}}%
      \expandafter\def\csname LTw\endcsname{\color{white}}%
      \expandafter\def\csname LTb\endcsname{\color{black}}%
      \expandafter\def\csname LTa\endcsname{\color{black}}%
      \expandafter\def\csname LT0\endcsname{\color{black}}%
      \expandafter\def\csname LT1\endcsname{\color{black}}%
      \expandafter\def\csname LT2\endcsname{\color{black}}%
      \expandafter\def\csname LT3\endcsname{\color{black}}%
      \expandafter\def\csname LT4\endcsname{\color{black}}%
      \expandafter\def\csname LT5\endcsname{\color{black}}%
      \expandafter\def\csname LT6\endcsname{\color{black}}%
      \expandafter\def\csname LT7\endcsname{\color{black}}%
      \expandafter\def\csname LT8\endcsname{\color{black}}%
    \fi
  \fi
    \setlength{\unitlength}{0.0500bp}%
    \ifx\gptboxheight\undefined%
      \newlength{\gptboxheight}%
      \newlength{\gptboxwidth}%
      \newsavebox{\gptboxtext}%
    \fi%
    \setlength{\fboxrule}{0.5pt}%
    \setlength{\fboxsep}{1pt}%
    \definecolor{tbcol}{rgb}{1,1,1}%
\begin{picture}(5260.00,4520.00)%
    \gplgaddtomacro\gplbacktext{%
      \colorrgb{0.00,0.00,0.00}%%
      \put(708,1018){\makebox(0,0)[r]{\strut{}$0.1$}}%
      \colorrgb{0.00,0.00,0.00}%%
      \put(708,2232){\makebox(0,0)[r]{\strut{}$1$}}%
      \colorrgb{0.00,0.00,0.00}%%
      \put(708,3446){\makebox(0,0)[r]{\strut{}$10$}}%
      \colorrgb{0.00,0.00,0.00}%%
      \put(882,448){\makebox(0,0){\strut{}$-0.3$}}%
      \colorrgb{0.00,0.00,0.00}%%
      \put(1505,448){\makebox(0,0){\strut{}$-0.2$}}%
      \colorrgb{0.00,0.00,0.00}%%
      \put(2127,448){\makebox(0,0){\strut{}$-0.1$}}%
      \colorrgb{0.00,0.00,0.00}%%
      \put(2749,448){\makebox(0,0){\strut{}$0$}}%
      \colorrgb{0.00,0.00,0.00}%%
      \put(3372,448){\makebox(0,0){\strut{}$0.1$}}%
      \colorrgb{0.00,0.00,0.00}%%
      \put(3994,448){\makebox(0,0){\strut{}$0.2$}}%
      \colorrgb{0.00,0.00,0.00}%%
      \put(4617,448){\makebox(0,0){\strut{}$0.3$}}%
    }%
    \gplgaddtomacro\gplfronttext{%
      \csname LTb\endcsname%%
      \put(186,2473){\rotatebox{-270}{\makebox(0,0){\strut{}Distribution $f(x)$}}}%
      \csname LTb\endcsname%%
      \put(2749,142){\makebox(0,0){\strut{}$x$}}%
    }%
    \gplbacktext
    \put(0,0){\includegraphics[width={263.00bp},height={226.00bp}]{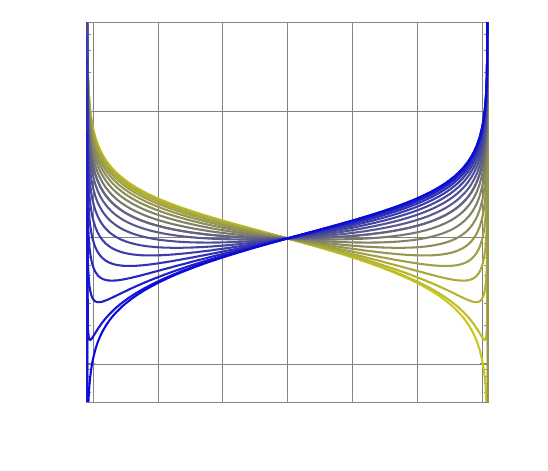}}%
    \gplfronttext
  \end{picture}%
\endgroup

%% file: main.bbl
\begin{thebibliography}{24}%
\makeatletter
\providecommand \@ifxundefined [1]{%
 \@ifx{#1\undefined}
}%
\providecommand \@ifnum [1]{%
 \ifnum #1\expandafter \@firstoftwo
 \else \expandafter \@secondoftwo
 \fi
}%
\providecommand \@ifx [1]{%
 \ifx #1\expandafter \@firstoftwo
 \else \expandafter \@secondoftwo
 \fi
}%
\providecommand \natexlab [1]{#1}%
\providecommand \enquote  [1]{``#1''}%
\providecommand \bibnamefont  [1]{#1}%
\providecommand \bibfnamefont [1]{#1}%
\providecommand \citenamefont [1]{#1}%
\providecommand \href@noop [0]{\@secondoftwo}%
\providecommand \href [0]{\begingroup \@sanitize@url \@href}%
\providecommand \@href[1]{\@@startlink{#1}\@@href}%
\providecommand \@@href[1]{\endgroup#1\@@endlink}%
\providecommand \@sanitize@url [0]{\catcode `\\12\catcode `\$12\catcode
  `\&12\catcode `\#12\catcode `\^12\catcode `\_12\catcode `\%12\relax}%
\providecommand \@@startlink[1]{}%
\providecommand \@@endlink[0]{}%
\providecommand \url  [0]{\begingroup\@sanitize@url \@url }%
\providecommand \@url [1]{\endgroup\@href {#1}{\urlprefix }}%
\providecommand \urlprefix  [0]{URL }%
\providecommand \Eprint [0]{\href }%
\providecommand \doibase [0]{https://doi.org/}%
\providecommand \selectlanguage [0]{\@gobble}%
\providecommand \bibinfo  [0]{\@secondoftwo}%
\providecommand \bibfield  [0]{\@secondoftwo}%
\providecommand \translation [1]{[#1]}%
\providecommand \BibitemOpen [0]{}%
\providecommand \bibitemStop [0]{}%
\providecommand \bibitemNoStop [0]{.\EOS\space}%
\providecommand \EOS [0]{\spacefactor3000\relax}%
\providecommand \BibitemShut  [1]{\csname bibitem#1\endcsname}%
\let\auto@bib@innerbib\@empty
%</preamble>
\bibitem [{\citenamefont
  {Kempe}(2003)}]{Kempe2003QuantumRandomWalksAnIntroductoryOverview}%
  \BibitemOpen
  \bibfield  {author} {\bibinfo {author} {\bibfnamefont {J.}~\bibnamefont
  {Kempe}},\ }\href {https://doi.org/10.1080/00107151031000110776} {\bibfield
  {journal} {\bibinfo  {journal} {Contemp. Phys.}\ }\textbf {\bibinfo {volume}
  {44}},\ \bibinfo {pages} {307} (\bibinfo {year} {2003})}\BibitemShut
  {NoStop}%
\bibitem [{\citenamefont
  {Schöning}(1999)}]{Schoening1999AProbabilisticAlgorithmForKSATAndConstraintSatisfactionProblems}%
  \BibitemOpen
  \bibfield  {author} {\bibinfo {author} {\bibfnamefont {U.}~\bibnamefont
  {Schöning}},\ }in\ \href {https://doi.org/10.1109/SFFCS.1999.814612} {\emph
  {\bibinfo {booktitle} {40th Annual Symposium on Foundations of Computer
  Science (Cat. No.99CB37039)}}}\ (\bibinfo {year} {1999})\ pp.\ \bibinfo
  {pages} {410--414}\BibitemShut {NoStop}%
\bibitem [{\citenamefont {Shenvi}\ \emph {et~al.}(2003)\citenamefont {Shenvi},
  \citenamefont {Kempe},\ and\ \citenamefont
  {Whaley}}]{Shenvi2003QuantumRandomWalkSearchAlgorithm}%
  \BibitemOpen
  \bibfield  {author} {\bibinfo {author} {\bibfnamefont {N.}~\bibnamefont
  {Shenvi}}, \bibinfo {author} {\bibfnamefont {J.}~\bibnamefont {Kempe}},\ and\
  \bibinfo {author} {\bibfnamefont {K.~B.}\ \bibnamefont {Whaley}},\ }\href
  {https://doi.org/10.1103/PhysRevA.67.052307} {\bibfield  {journal} {\bibinfo
  {journal} {Phys. Rev. A}\ }\textbf {\bibinfo {volume} {67}},\ \bibinfo
  {pages} {052307} (\bibinfo {year} {2003})}\BibitemShut {NoStop}%
\bibitem [{\citenamefont
  {Ambainis}(2003)}]{Ambainis2003QuantumWalksAndTheirAlgorithmicApplications}%
  \BibitemOpen
  \bibfield  {author} {\bibinfo {author} {\bibfnamefont {A.}~\bibnamefont
  {Ambainis}},\ }\href {https://doi.org/10.1142/S0219749903000383} {\bibfield
  {journal} {\bibinfo  {journal} {Int. J. Quantum Inf.}\ }\textbf {\bibinfo
  {volume} {01}},\ \bibinfo {pages} {507} (\bibinfo {year} {2003})}\BibitemShut
  {NoStop}%
\bibitem [{\citenamefont {Magniez}\ \emph {et~al.}(2007)\citenamefont
  {Magniez}, \citenamefont {Nayak}, \citenamefont {Roland},\ and\ \citenamefont
  {Santha}}]{Magniez2007SearchViaQuantumWalk}%
  \BibitemOpen
  \bibfield  {author} {\bibinfo {author} {\bibfnamefont {F.}~\bibnamefont
  {Magniez}}, \bibinfo {author} {\bibfnamefont {A.}~\bibnamefont {Nayak}},
  \bibinfo {author} {\bibfnamefont {J.}~\bibnamefont {Roland}},\ and\ \bibinfo
  {author} {\bibfnamefont {M.}~\bibnamefont {Santha}},\ }in\ \href
  {https://doi.org/10.1145/1250790.1250874} {\emph {\bibinfo {booktitle}
  {Proceedings of the Thirty-Ninth Annual ACM Symposium on Theory of
  Computing}}},\ \bibinfo {series and number} {STOC '07}\ (\bibinfo
  {publisher} {Association for Computing Machinery},\ \bibinfo {address} {New
  York, NY, USA},\ \bibinfo {year} {2007})\ p.\ \bibinfo {pages}
  {575–584}\BibitemShut {NoStop}%
\bibitem [{\citenamefont {Kitagawa}\ \emph {et~al.}(2010)\citenamefont
  {Kitagawa}, \citenamefont {Rudner}, \citenamefont {Berg},\ and\ \citenamefont
  {Demler}}]{Kitagawa2010ExploringTopologicalPhasesWithQuantumWalks}%
  \BibitemOpen
  \bibfield  {author} {\bibinfo {author} {\bibfnamefont {T.}~\bibnamefont
  {Kitagawa}}, \bibinfo {author} {\bibfnamefont {M.~S.}\ \bibnamefont
  {Rudner}}, \bibinfo {author} {\bibfnamefont {E.}~\bibnamefont {Berg}},\ and\
  \bibinfo {author} {\bibfnamefont {E.}~\bibnamefont {Demler}},\ }\href
  {https://doi.org/10.1103/PhysRevA.82.033429} {\bibfield  {journal} {\bibinfo
  {journal} {Phys. Rev. A}\ }\textbf {\bibinfo {volume} {82}},\ \bibinfo
  {pages} {033429} (\bibinfo {year} {2010})}\BibitemShut {NoStop}%
\bibitem [{\citenamefont {Asb\'oth}\ \emph {et~al.}(2014)\citenamefont
  {Asb\'oth}, \citenamefont {Tarasinski},\ and\ \citenamefont
  {Delplace}}]{Asboth2014ChiralSymmetryAndBulkBoundaryCorrespondenceInPeriodicallyDrivenOneDimensionalSystems}%
  \BibitemOpen
  \bibfield  {author} {\bibinfo {author} {\bibfnamefont {J.~K.}\ \bibnamefont
  {Asb\'oth}}, \bibinfo {author} {\bibfnamefont {B.}~\bibnamefont
  {Tarasinski}},\ and\ \bibinfo {author} {\bibfnamefont {P.}~\bibnamefont
  {Delplace}},\ }\href {https://doi.org/10.1103/PhysRevB.90.125143} {\bibfield
  {journal} {\bibinfo  {journal} {Phys. Rev. B}\ }\textbf {\bibinfo {volume}
  {90}},\ \bibinfo {pages} {125143} (\bibinfo {year} {2014})}\BibitemShut
  {NoStop}%
\bibitem [{\citenamefont {Cedzich}\ \emph {et~al.}(2018)\citenamefont
  {Cedzich}, \citenamefont {Geib}, \citenamefont {Gr{\"u}nbaum}, \citenamefont
  {Stahl}, \citenamefont {Vel{\'a}zquez}, \citenamefont {Werner},\ and\
  \citenamefont
  {Werner}}]{Cedzich2018TheTopologicalClassificationOfOneDimensionalSymmetricQuantumWalks}%
  \BibitemOpen
  \bibfield  {author} {\bibinfo {author} {\bibfnamefont {C.}~\bibnamefont
  {Cedzich}}, \bibinfo {author} {\bibfnamefont {T.}~\bibnamefont {Geib}},
  \bibinfo {author} {\bibfnamefont {F.~A.}\ \bibnamefont {Gr{\"u}nbaum}},
  \bibinfo {author} {\bibfnamefont {C.}~\bibnamefont {Stahl}}, \bibinfo
  {author} {\bibfnamefont {L.}~\bibnamefont {Vel{\'a}zquez}}, \bibinfo {author}
  {\bibfnamefont {A.~H.}\ \bibnamefont {Werner}},\ and\ \bibinfo {author}
  {\bibfnamefont {R.~F.}\ \bibnamefont {Werner}},\ }\href
  {https://doi.org/10.1007/s00023-017-0630-x} {\bibfield  {journal} {\bibinfo
  {journal} {Ann. Henri Poincar{\'e}}\ }\textbf {\bibinfo {volume} {19}},\
  \bibinfo {pages} {325} (\bibinfo {year} {2018})}\BibitemShut {NoStop}%
\bibitem [{\citenamefont {\ifmmode \check{S}\else
  \v{S}\fi{}tefa\ifmmode~\check{n}\else \v{n}\fi{}\'ak}\ \emph
  {et~al.}(2008)\citenamefont {\ifmmode \check{S}\else
  \v{S}\fi{}tefa\ifmmode~\check{n}\else \v{n}\fi{}\'ak}, \citenamefont {Kiss},\
  and\ \citenamefont
  {Jex}}]{Stefanak2008RecurrencePropertiesOfUnbiasedCoinedQuantumWalksOnInfiniteDimensionalLattices}%
  \BibitemOpen
  \bibfield  {author} {\bibinfo {author} {\bibfnamefont {M.}~\bibnamefont
  {\ifmmode \check{S}\else \v{S}\fi{}tefa\ifmmode~\check{n}\else
  \v{n}\fi{}\'ak}}, \bibinfo {author} {\bibfnamefont {T.}~\bibnamefont
  {Kiss}},\ and\ \bibinfo {author} {\bibfnamefont {I.}~\bibnamefont {Jex}},\
  }\href {https://doi.org/10.1103/PhysRevA.78.032306} {\bibfield  {journal}
  {\bibinfo  {journal} {Phys. Rev. A}\ }\textbf {\bibinfo {volume} {78}},\
  \bibinfo {pages} {032306} (\bibinfo {year} {2008})}\BibitemShut {NoStop}%
\bibitem [{\citenamefont {Mohseni}\ \emph {et~al.}(2008)\citenamefont
  {Mohseni}, \citenamefont {Rebentrost}, \citenamefont {Lloyd},\ and\
  \citenamefont
  {Aspuru-Guzik}}]{Mohseni2008EnvironmentAssistedQuantumWalksInPhotosyntheticEnergyTransfer}%
  \BibitemOpen
  \bibfield  {author} {\bibinfo {author} {\bibfnamefont {M.}~\bibnamefont
  {Mohseni}}, \bibinfo {author} {\bibfnamefont {P.}~\bibnamefont {Rebentrost}},
  \bibinfo {author} {\bibfnamefont {S.}~\bibnamefont {Lloyd}},\ and\ \bibinfo
  {author} {\bibfnamefont {A.}~\bibnamefont {Aspuru-Guzik}},\ }\href
  {https://doi.org/10.1063/1.3002335} {\bibfield  {journal} {\bibinfo
  {journal} {J. Chem. Phys.}\ }\textbf {\bibinfo {volume} {129}},\ \bibinfo
  {pages} {174106} (\bibinfo {year} {2008})}\BibitemShut {NoStop}%
\bibitem [{\citenamefont {Oliveira}\ \emph {et~al.}(2006)\citenamefont
  {Oliveira}, \citenamefont {Portugal},\ and\ \citenamefont
  {Donangelo}}]{Oliveira2006DecoherenceInTwoDimensionalQuantumWalks}%
  \BibitemOpen
  \bibfield  {author} {\bibinfo {author} {\bibfnamefont {A.~C.}\ \bibnamefont
  {Oliveira}}, \bibinfo {author} {\bibfnamefont {R.}~\bibnamefont {Portugal}},\
  and\ \bibinfo {author} {\bibfnamefont {R.}~\bibnamefont {Donangelo}},\ }\href
  {https://doi.org/10.1103/PhysRevA.74.012312} {\bibfield  {journal} {\bibinfo
  {journal} {Phys. Rev. A}\ }\textbf {\bibinfo {volume} {74}},\ \bibinfo
  {pages} {012312} (\bibinfo {year} {2006})}\BibitemShut {NoStop}%
\bibitem [{\citenamefont
  {Chandrashekar}(2010)}]{Chandrashekar2010DiscreteTimeQuantumWalkDynamicsAndApplications}%
  \BibitemOpen
  \bibfield  {author} {\bibinfo {author} {\bibfnamefont {C.~M.}\ \bibnamefont
  {Chandrashekar}},\ }\href@noop {} {\bibinfo {title} {{Discrete-Time Quantum
  Walk - Dynamics and Applications}}} (\bibinfo {year} {2010}),\ \Eprint
  {https://arxiv.org/abs/1001.5326} {arXiv:1001.5326 [quant-ph]} \BibitemShut
  {NoStop}%
\bibitem [{\citenamefont {Farhi}\ and\ \citenamefont
  {Gutmann}(1998)}]{Farhi1998QuantumComputationAndDecisionTrees}%
  \BibitemOpen
  \bibfield  {author} {\bibinfo {author} {\bibfnamefont {E.}~\bibnamefont
  {Farhi}}\ and\ \bibinfo {author} {\bibfnamefont {S.}~\bibnamefont
  {Gutmann}},\ }\href {https://doi.org/10.1103/PhysRevA.58.915} {\bibfield
  {journal} {\bibinfo  {journal} {Phys. Rev. A}\ }\textbf {\bibinfo {volume}
  {58}},\ \bibinfo {pages} {915} (\bibinfo {year} {1998})}\BibitemShut
  {NoStop}%
\bibitem [{\citenamefont {Aharonov}\ \emph {et~al.}(1993)\citenamefont
  {Aharonov}, \citenamefont {Davidovich},\ and\ \citenamefont
  {Zagury}}]{Aharonov1993QuantumRandomWalks}%
  \BibitemOpen
  \bibfield  {author} {\bibinfo {author} {\bibfnamefont {Y.}~\bibnamefont
  {Aharonov}}, \bibinfo {author} {\bibfnamefont {L.}~\bibnamefont
  {Davidovich}},\ and\ \bibinfo {author} {\bibfnamefont {N.}~\bibnamefont
  {Zagury}},\ }\href {https://doi.org/10.1103/PhysRevA.48.1687} {\bibfield
  {journal} {\bibinfo  {journal} {Phys. Rev. A}\ }\textbf {\bibinfo {volume}
  {48}},\ \bibinfo {pages} {1687} (\bibinfo {year} {1993})}\BibitemShut
  {NoStop}%
\bibitem [{\citenamefont {Aharonov}\ \emph {et~al.}(2001)\citenamefont
  {Aharonov}, \citenamefont {Ambainis}, \citenamefont {Kempe},\ and\
  \citenamefont {Vazirani}}]{Aharonov2001QuantumWalksOnGraphs}%
  \BibitemOpen
  \bibfield  {author} {\bibinfo {author} {\bibfnamefont {D.}~\bibnamefont
  {Aharonov}}, \bibinfo {author} {\bibfnamefont {A.}~\bibnamefont {Ambainis}},
  \bibinfo {author} {\bibfnamefont {J.}~\bibnamefont {Kempe}},\ and\ \bibinfo
  {author} {\bibfnamefont {U.}~\bibnamefont {Vazirani}},\ }in\ \href
  {https://doi.org/10.1145/380752.380758} {\emph {\bibinfo {booktitle}
  {Proceedings of the Thirty-Third Annual ACM Symposium on Theory of
  Computing}}},\ \bibinfo {series and number} {STOC '01}\ (\bibinfo
  {publisher} {Association for Computing Machinery},\ \bibinfo {address} {New
  York, NY, USA},\ \bibinfo {year} {2001})\ p.\ \bibinfo {pages}
  {50–59}\BibitemShut {NoStop}%
\bibitem [{\citenamefont {Ambainis}\ \emph {et~al.}(2001)\citenamefont
  {Ambainis}, \citenamefont {Bach}, \citenamefont {Nayak}, \citenamefont
  {Vishwanath},\ and\ \citenamefont
  {Watrous}}]{Ambainis2001OneDimensionalQuantumWalks}%
  \BibitemOpen
  \bibfield  {author} {\bibinfo {author} {\bibfnamefont {A.}~\bibnamefont
  {Ambainis}}, \bibinfo {author} {\bibfnamefont {E.}~\bibnamefont {Bach}},
  \bibinfo {author} {\bibfnamefont {A.}~\bibnamefont {Nayak}}, \bibinfo
  {author} {\bibfnamefont {A.}~\bibnamefont {Vishwanath}},\ and\ \bibinfo
  {author} {\bibfnamefont {J.}~\bibnamefont {Watrous}},\ }in\ \href
  {https://doi.org/10.1145/380752.380757} {\emph {\bibinfo {booktitle}
  {Proceedings of the Thirty-Third Annual ACM Symposium on Theory of
  Computing}}},\ \bibinfo {series and number} {STOC '01}\ (\bibinfo
  {publisher} {Association for Computing Machinery},\ \bibinfo {address} {New
  York, NY, USA},\ \bibinfo {year} {2001})\ p.\ \bibinfo {pages}
  {37–49}\BibitemShut {NoStop}%
\bibitem [{\citenamefont
  {Feller}(1991)}]{Feller1991AnIntroductionToProbabilityTheoryAndItsApplications}%
  \BibitemOpen
  \bibfield  {author} {\bibinfo {author} {\bibfnamefont {W.}~\bibnamefont
  {Feller}},\ }\href@noop {} {\emph {\bibinfo {title} {An introduction to
  probability theory and its applications}}},\ Vol.~\bibinfo {volume} {2}\
  (\bibinfo  {publisher} {John Wiley \& Sons},\ \bibinfo {year}
  {1991})\BibitemShut {NoStop}%
\bibitem [{\citenamefont
  {Konno}(2002)}]{Konno2002QuantumRandomWalksInOneDimension}%
  \BibitemOpen
  \bibfield  {author} {\bibinfo {author} {\bibfnamefont {N.}~\bibnamefont
  {Konno}},\ }\href {https://doi.org/10.1023/A:1023413713008} {\bibfield
  {journal} {\bibinfo  {journal} {Quantum Inf. Process.}\ }\textbf {\bibinfo
  {volume} {1}},\ \bibinfo {pages} {345} (\bibinfo {year} {2002})}\BibitemShut
  {NoStop}%
\bibitem [{\citenamefont
  {Goldstein}(1951)}]{Goldstein1951OnDiffusionByDiscontinuousMovementsAndOnTheTelegraphEquation}%
  \BibitemOpen
  \bibfield  {author} {\bibinfo {author} {\bibfnamefont {S.}~\bibnamefont
  {Goldstein}},\ }\href {https://doi.org/10.1093/qjmam/4.2.129} {\bibfield
  {journal} {\bibinfo  {journal} {Q. J. Mech. Appl. Math.}\ }\textbf {\bibinfo
  {volume} {4}},\ \bibinfo {pages} {129} (\bibinfo {year} {1951})}\BibitemShut
  {NoStop}%
\bibitem [{\citenamefont {Gillis}(1995)}]{Gillis1955CorrelatedRandomWalk}%
  \BibitemOpen
  \bibfield  {author} {\bibinfo {author} {\bibfnamefont {J.}~\bibnamefont
  {Gillis}},\ }\href {https://doi.org/10.1017/S0305004100030711} {\bibfield
  {journal} {\bibinfo  {journal} {Math. Proc. Camb. Philos. Soc.}\ }\textbf
  {\bibinfo {volume} {51}},\ \bibinfo {pages} {639} (\bibinfo {year}
  {1995})}\BibitemShut {NoStop}%
\bibitem [{\citenamefont {Jayakody}\ and\ \citenamefont
  {Cohen}(2023)}]{Jayakody2023ClosedFormExpressionsForTheProbabilityDistributionOfQuantumWalkOnALine}%
  \BibitemOpen
  \bibfield  {author} {\bibinfo {author} {\bibfnamefont {M.~N.}\ \bibnamefont
  {Jayakody}}\ and\ \bibinfo {author} {\bibfnamefont {E.}~\bibnamefont
  {Cohen}},\ }\href {https://doi.org/10.1140/epjd/s10053-023-00780-9}
  {\bibfield  {journal} {\bibinfo  {journal} {Eur. Phys. J. D}\ }\textbf
  {\bibinfo {volume} {77}},\ \bibinfo {pages} {193} (\bibinfo {year}
  {2023})}\BibitemShut {NoStop}%
\bibitem [{\citenamefont
  {Konno}(2005)}]{Konno2005ANewTypeOfLimitTheoremsForTheOneDimensionalQuantumRandomWalk}%
  \BibitemOpen
  \bibfield  {author} {\bibinfo {author} {\bibfnamefont {N.}~\bibnamefont
  {Konno}},\ }\href {https://doi.org/10.2969/jmsj/1150287309} {\bibfield
  {journal} {\bibinfo  {journal} {J. Math. Soc. Japan}\ }\textbf {\bibinfo
  {volume} {57}},\ \bibinfo {pages} {1179} (\bibinfo {year}
  {2005})}\BibitemShut {NoStop}%
\bibitem [{\citenamefont {Nayak}\ and\ \citenamefont
  {Vishwanath}(2000)}]{Nayak2000QuantumWalkOnTheLine}%
  \BibitemOpen
  \bibfield  {author} {\bibinfo {author} {\bibfnamefont {A.}~\bibnamefont
  {Nayak}}\ and\ \bibinfo {author} {\bibfnamefont {A.}~\bibnamefont
  {Vishwanath}},\ }\href@noop {} {\bibinfo {title} {{Quantum Walk on the
  Line}}} (\bibinfo {year} {2000}),\ \Eprint
  {https://arxiv.org/abs/quant-ph/0010117} {arXiv:quant-ph/0010117}
  \BibitemShut {NoStop}%
\bibitem [{\citenamefont {Taher}\ \emph {et~al.}(2006)\citenamefont {Taher},
  \citenamefont {Mouline},\ and\ \citenamefont
  {Rachidi}}]{Taher2006FibonacciHornerDecompositionOfTheMatrixExponentialAndTheFundamentalSystemOfSolutions}%
  \BibitemOpen
  \bibfield  {author} {\bibinfo {author} {\bibfnamefont {R.}~\bibnamefont
  {Taher}}, \bibinfo {author} {\bibfnamefont {M.}~\bibnamefont {Mouline}},\
  and\ \bibinfo {author} {\bibfnamefont {M.}~\bibnamefont {Rachidi}},\ }\href
  {https://doi.org/10.13001/1081-3810.1228} {\bibfield  {journal} {\bibinfo
  {journal} {Electron. J. Linear Algebra}\ }\textbf {\bibinfo {volume} {15}}
  (\bibinfo {year} {2006})}\BibitemShut {NoStop}%
\end{thebibliography}%
